\documentclass[12pt,journal,final,letterpaper,onecolumn]{article}
\usepackage[top=1in, bottom=1in, left=1in, right=1in]{geometry}

\usepackage{graphicx} 
\usepackage{subfigure}
\usepackage{algorithm}
\usepackage{algorithmic}
\usepackage{hyperref}
\usepackage{amsmath,  amssymb, amsthm}
\usepackage[comma,authoryear]{natbib}

\usepackage{color}
\usepackage{booktabs}

\usepackage{setspace}

\newcommand{\beq}{\vspace{0mm}\begin{equation}}
\newcommand{\eeq}{\vspace{0mm}\end{equation}}
\newcommand{\beqs}{\vspace{0mm}\begin{eqnarray}}
\newcommand{\eeqs}{\vspace{0mm}\end{eqnarray}}
\newcommand{\barr}{\begin{array}}
\newcommand{\earr}{\end{array}}

\newcommand{\Lmat}[0]{{{\bf L}}}

\newcommand{\Nmat}[0]{{{\bf N}}}

\newcommand{\nv}[0]{{\boldsymbol{n}}}

\newcommand{\pv}[0]{{\boldsymbol{p}}}

\newcommand{\rv}{\boldsymbol{r}}

\newcommand{\cdotv}{\boldsymbol{\cdot}}

\newcommand{\thetav}{\boldsymbol{\theta}}

\newcommand{\E}{\mathbb{E}}

\usepackage{natbib} 

\renewcommand\footnotemark{}

\begin{document}
\title{Priors for Random Count Matrices Derived from a Family of Negative Binomial Processes}
\author{Mingyuan~Zhou$^\dagger$$^\ddag$,  
Oscar Hernan 
Madrid Padilla$^\ddag$, 
and James~G.~Scott$^\dagger$$^\ddag$\\ 
The University of Texas at Austin
%
\thanks{ 
The authors are with the $^\dagger$Department of Information, Risk, and Operations Management and $^\ddag$Department of Statistics and Data Sciences, the University of Texas at Austin, Austin, TX 78712, USA. \emph{Address for correspondence}: 2110 Speedway Stop B6500, Austin, TX 78712, USA.
\emph{Email:} \texttt{mingyuan.zhou@mccombs.utexas.edu}.
}
}


\begin{spacing}{1.05}

\maketitle

\begin{abstract}

We define a family of probability distributions for random count matrices with a potentially unbounded number of rows and columns.  The three distributions we consider are derived from the gamma-Poisson, gamma-negative binomial, and beta-negative binomial processes, which we refer to generically as a family of negative-binomial processes.   Because the models lead to closed-form update equations within the context of a Gibbs sampler, they are natural candidates for nonparametric Bayesian priors over count matrices.  A key aspect of our analysis is the recognition that, although the random count matrices within the family are defined by a row-wise construction, their columns can be shown to be independent and identically distributed.  This fact is used to derive explicit formulas for drawing all the columns at once.  Moreover, by analyzing these matrices' combinatorial structure, we describe how to sequentially construct a column-i.i.d.~random count matrix one row at a time, and derive the predictive distribution of a new row count vector with previously unseen features.  We describe the similarities and differences between the three priors, and argue that the greater flexibility of the gamma- and beta- negative binomial processes---especially their ability to model over-dispersed, heavy-tailed count data---makes these well suited to a wide variety of real-world applications.  As an example of our framework, we construct a naive-Bayes text classifier to categorize a count vector to one of several existing random count matrices of different categories.  The classifier supports an unbounded number of features, and unlike most existing methods, it  does not require  a predefined finite vocabulary to be shared by all the categories, and needs neither feature selection nor parameter tuning.  Both the gamma- and beta- negative binomial processes are shown to significantly outperform the gamma-Poisson process when applied to document categorization, with comparable performance to other state-of-the-art supervised text classification algorithms.

\end{abstract} 

\end{spacing}

\newpage

\begin{spacing}{1.5}

\section{Introduction}
\label{sec:intro}
\subsection{Models for count matrices}

The need to model a random count matrix arises in many settings, from linguistics to marketing to ecology.  For example, in text analysis, we often observe a document-term matrix, whose rows record how many times word $k$ appeared in a given document.  In a biodiversity study, we may observe a site-species matrix, where each row records the number of times species $k$ was observed at a given site.  Similar applications arise in a wide variety of fields; for examples, see \citet{Cameron1998,Chib1998,CannyGaP,DCA,WinkelmannCount,InfGaP}, and \citet{BNBP_PFA_AISTATS2012}.

Nonparametric Bayesian analysis provides a natural setting in which to study random matrices, especially those with no natural upper bound on the number of rows or columns.  Yet while there is a wide selection of nonparametric  Bayesian models for random count vectors and random binary matrices, prior distributions over random count matrices are relatively underdeveloped.  Moreover, a major conceptual problem in modeling a random count matrix arises when new rows are added sequentially.  For example, as new documents are collected and processed in text analysis, each new document (represented by a new row of the matrix) may contain previously unseen words (features).  This requires that new columns be added to the existing count matrix.  But it is not obvious how to define the predictive distribution of this new row of a random count matrix, if the row contains previously unseen features.  This is especially important in natural language processing, where a common application is to build a naive Bayes model for classifying new documents.  Without having a predictive distribution that accounts for new features, one must often use a predetermined vocabulary and simply ignore the previously unseen terms appearing in a new document.    

We directly address these issues by investigating a family of nonparametric Bayesian priors for random count matrices constructed from stochastic processes: the gamma-Poisson process, the gamma-negative binomial process (GNBP), and the beta-negative binomial process (BNBP).  We show that all these processes lead to random count matrices with independent and identically distributed (i.i.d.) columns, which can be constructed by drawing all the columns at once, or by adding one row at a time.  In addition, we show the gamma-Poisson process, and for special cases of the GNBP and BNBP with common row-wise parameters, the generated random count matrices are exchangeable in both rows and columns.

Our derivation exactly marginalizes out the underlying stochastic processes to arrive at a probability mass function (PMF) for a column-i.i.d. random count matrix.   In contrast to existing techniques that take the infinite limit of a finite-dimensional model, this novel procedure allows for the construction and analysis of much more flexible nonparametric priors for random matrices, and highlights certain model properties that are not evident from the finite-model limit.   The argument relies upon a novel combinatorial analysis for calculating the number of ways to map a column-i.i.d. random count matrix to a structured random count matrix whose columns are ordered in a certain manner.  This is a key step in deriving the predictive distribution of a new random count vector under a random count matrix.

As an application of our proposed framework, we construct a naive-Bayes text classification model.  The approach does not require a predefined list of terms (features), and naturally accounts for documents with previously unseen terms.  This also implies that random count matrices of different categories can be updated, analyzed, and tested completely in parallel. Moreover, the algorithm requires neither feature selection nor parameter tuning. Following \citet{crammer2012confidence}, the algorithm may also be conveniently extended to an online learning setting.   Empirical results suggest that both the proposed GNBP and BNBP models lead to substantially better out-of-sample classification performance, versus both the gamma-Poisson model and the multinomial model with Laplace smoothing. They also clearly outperform the text classification algorithms that first  learn lower-dimensional feature vectors for documents and then  train a multi-class classifier, and have comparable performance to the state-of-the-art discriminatively trained  text classification algorithms, whose  features need to be carefully constructed and parameters  carefully selected. 

\subsection{Connections with existing work}

Our paper is in the spirit of existing work on nonparametric Bayesian priors for random count vectors and random binary matrices.  To model a random count vector, one may use the Chinese restaurant process, or any one of many other stochastic processes characterized by exchangeable partition probability functions (EPPFs) or sample-size dependent EPPFs; see, for example, \citet{PolyaUrn,csp,BeyondDP}, and 
\citet{gNBP2014}.  Likewise, to model a random binary matrix, one may use the Indian buffet process \citep{IBP,SBP}.  These well-studied nonparametric Bayesian priors, however, are not directly useful for describing random count matrices. To address this gap, we investigate a family of nonparametric Bayesian priors for random count matrices, each based on a  previously proposed stochastic process that has not been thoroughly studied: the gamma-Poisson process \citep{lo1982bayesian,InfGaP}, the gamma-negative binomial process, or GNBP \citep{NBP2012};
and the beta-negative binomial process, or BNBP \citep{BNBP_PFA_AISTATS2012,NBPJordan}. 

All three models can be derived as the marginal distribution of a suitably defined stochastic process with respect to a traditional sampling model for integer-valued counts.  This parallels the construction of the models for count vectors or binary matrices mentioned previously.  For example, the Chinese restaurant process describes  a random count vector as the marginal of the Dirichlet process \citep{ferguson73} under multinomial sampling.  Likewise, the Indian buffet process describes a random  binary matrix as the marginal of the beta process \citep{Hjort} under Bernoulli sampling \citep{JordanBP}.  Similarly, we present the negative binomial process as the marginal of the gamma process under Poisson sampling, the
GNBP as the marginal of the  gamma process under negative binomial sampling, and the BNBP as the marginal of the beta process under negative binomial sampling.

The remainder of the paper is organized as follows. After some preliminary definitions and notation,  we introduce in Section \ref{sec:random_count_matrix} three distinct nonparametric Bayesian priors for random count matrices. In Section \ref{sec:experiments}, we  construct nonparametric Bayesian naive Bayes classifiers to classifier a count vector to one of several existing count matrices and demonstrate their use in document categorization. The details for deriving the random count matrix priors from their underlying hierarchical stochastic processes are provided in the Supplementary Material.

\subsection{Notation and preliminaries}

\label{sec:prelims}
\paragraph{Stochastic processes.} A gamma process \citep{ferguson73} 
$G\sim\Gamma{\mbox{P}}(G_0,1/c)$ on the product space $\mathbb{R}_+\times \Omega$, where $\mathbb{R}_+=\{x:x>0\}$, is defined by two parameters: a  finite and continuous base measure $G_0$ over a complete separable metric  space $\Omega$, and a scale $1/c$, such that $G(A)\sim\mbox{Gamma}(G_0(A),1/c)$ for each $A\subset \Omega$.  The L\'evy measure of the gamma process is $\nu(drd\omega)=r^{-1}e^{-cr}dr G_0(d\omega)$.  Although the L\'evy measure integrates to infinity, $\int _{\mathbb{R}_+\times \Omega} 
\min\{r,1\}\nu(drd\omega)$ is finite,  and therefore a draw from the gamma process 
$G\sim\Gamma\mbox{P}(G_0,1/c)$ can be represented as the countably infinite sum
$G=\sum_{k=1}^\infty r_k\delta_{\omega_k},~\omega_k\sim g_0,$ where $\gamma_0=G_0(\Omega)$ is the mass parameter and $g_0(d\omega)=G_0(d\omega)/\gamma_0$ is the base distribution.

A beta process \citep{Hjort} $B\sim\mbox{BP}(c,B_0)$ 
on the product space $[0,1]\times \Omega$, is also defined by two parameters: a finite and continuous base measure $B_0$ over a complete separable metric  space $\Omega$, and a concentration parameter $c>0$. The L\'evy measure  of the beta process in this paper is defined  as
\begin{equation}
\label{eq:BP_Levy}
\nu(dpd\omega)=p^{-1}(1-p)^{c-1} dp B_0(d\omega) \, .
\end{equation}
As $\int_{[0,1]\times \Omega}\nu(dpd\omega)=\infty$ and $\int_{[0,1]\times \Omega}\min\{p,1\}\nu(dpd\omega)<\infty$, 
a draw from  $B\sim\mbox{BP}(c,B_0)$ can be represented   as $B=\sum_{k=1}^\infty p_k\delta_{\omega_k},~\omega_k\sim g_0,$ where $\gamma_0=B_0(\Omega)$ is the mass parameter and $g_0(d\omega)=B_0(d\omega)/\gamma_0$ is the base distribution.

\paragraph{Random count matrices.}

A random count matrix is denoted generically  by $\Nmat_J \in \mathbb{Z}^{J\times K_J}$, $\mathbb{Z} = \{0, 1, \ldots\}$, where the $J$ rows of $\Nmat_J$ correspond to the $J$ samples or cases, and the $K_J$ columns to features that have been observed at least once across all rows.  Throughout the paper, we will refer to count matrices constructed sequentially by row, for which we require a consistent notation.  Suppose that a new case is observed; we use $\Nmat^{+}_{J+1}$  to refer to the new part introduced to the matrix $\Nmat_J$ by adding row $(J+1)$.  
Similarly, we use $K^+_{J+1}$ to denote the number of new columns introduced by adding row $(J+1)$, meaning that $K_{J+1}:=K_J+K^+_{J+1}$; $\nv_{:k}$ to indicate the 
count vector corresponding to column $k$ of the matrix; and $n_{\cdot k} = \sum_{j=1}^{K_J} \nv_{:k} $ to denote the total number of counts of feature $k$ across all rows. One may think of $\Nmat^+_{J+1}$ as the combination of two submatrices: a row of $K_J$ counts appended below $\Nmat_J$, and then a $(J+1) \times K^+_{J+1}$ submatrix, whose first $J$ rows are entirely zero, and whose $K^+_{J+1}$ columns are inserted into random locations among original columns with their relative orders preserved.

Our convention is that a prior for a random count matrix is named by the stochastic process used to generate each of its rows.  In this paper, we study three hierarchical  stochastic processes, all in the family of negative binomial processes. Each such stochastic process is defined by the prior for an almost-surely discrete random measure, together with a sampling model for generating counts.  We denote the distribution of such a matrix as $\Nmat \sim \mbox{ProcessM}(\thetav)$, where ``Process'' is the name of the underlying hierarchical stochastic process, ``M'' stands for matrix, and $\thetav$ encodes the parameters of the process.

For example, to construct a gamma-Poisson or negative binomial process random count matrix, $\Nmat_J \sim \mbox{NBPM}(\gamma_0, c)$,  we draw a random measure $G \sim \Gamma \mbox{P}(G_0, 1/c)$ from a gamma process.  Then for each row of the matrix, we independently draw $X_j \mid G \sim \mbox{PP}(G)$: a Poisson process such that $X_j(A) \sim \mbox{Pois}[G(A)]$ for all $A \subset \Omega$.  As $G=\sum_{k=1}^\infty r_k\delta_{\omega_k}$ is atomic, we have $X_j=\sum_{k=1}^\infty n_{jk}\delta_{\omega_k},~n_{jk}\sim\mbox{Pois}(r_k)$. Although $\{X_j\}_{1,J}$ contains countably many atoms, we will show in later sections that only a finite number of them have nonzero counts. The count matrix $\Nmat_J$ is constructed by organizing all the nonzero column count vectors, $\{\nv_{:k }\}_{k:n_{\cdotv k}>0}$, in an arbitrary order into a random  count matrix. 
Thus the statistical features we care about, such as words or species, are identified with the atoms of the underlying random measure.

\paragraph{Some important distributions.}

The notation  $u\sim\mbox{Log}(p)$ denotes a random variable having a logarithmic distribution \citep{LogPoisNB} 
with PMF
$$
f_U(u \mid p) = \frac{1}{-\ln(1-p)}\frac{p^u}{u} \quad \mbox{for  } u \in \{1,2,\ldots\} \, .
$$ 

A related distribution, called the sum-logarithmic, is defined as follows.  Let $u_t\sim\mbox{Log}(p)$, and let $n=\sum_{t=1}^{l}u_{t}$.  The marginal distribution of $n$ is a sum-logarithmic distribution \citep{NBP2012}, expressed as $n\sim\mbox{SumLog}(l,p)$, with PMF
$$
f_{N}(n \mid l,p) = \frac{ p^n l! \ |s(n,l)|}{n! \ [-\ln(1-p)]^l} \, ,
$$
where $|s(n,l)|$ are unsigned Stirling numbers of the first kind.  These are related to gamma functions by
\beq\label{eq:gammafunctions}
\frac{\Gamma(n+r)}{\Gamma(r)}=\sum_{l=0}^n |s(n,l)|r^l \, .
\eeq

The joint distribution of $n\sim\mbox{SumLog}(l,p)$ and $l\sim\mbox{Pois}[-r\ln(1-p)]$ is described as the Poisson-logarithmic bivariate distribution in \citet{NBP2012}, with PMF
\begin{equation}
\label{eqn:NBcompoundPoisson}
f_{N,L}(n,l \mid r,p )=\frac{|s(n,l)|r^l}{n!} p^{n} (1-p)^{r} \, .
\end{equation}
The marginalization of $l$ from this compound Poisson representation leads to the negative binomial distribution $n\sim\mbox{NB}(r,p)$, 
with PMF 
 $$
f_{N}(n \mid r,p )=\frac{\Gamma(n+r)}{n!\Gamma(r)} p^{n} (1-p)^{r} \, .
$$

We describe 
in 
the Supplementary Material
several other useful  distributions, including the logarithmic mixed sum-logarithmic (LogLog), the negative binomial mixed sum-logarithmic, the gamma-negative binomial (GNB),  the beta-negative binomial (BNB), the digamma distribution, and the logbeta distributions.

\section{Nonparametric Priors for Random Count Matrices}\label{sec:random_count_matrix}

In this section, we introduce three nonparametric Bayesian priors for random count matrices; for the gamma-Poisson process,  we describe in detail its PMF,  row- and column-wise construction, and some other basic properties; and for the GNBP and BNBP, we present their PMFs and defer other details to the Supplementary Material. We then describe the predictive distribution of a new row count vector under a random count matrix, and highlight some important differences among the three priors.  Although results here are quoted without proof, and the detailed construction is deferred to the Supplementary Material, the basic manner of argument in each case is similar.  Our goal is to marginalize out the infinite-dimensional  random measure to obtain the unconditional PMF of the random count matrix $\Nmat_J\in\mathbb{Z}^{J\times K_J}$, where $\mathbb{Z}=\{0,1,\ldots\}$.  We are able to do so by separating the absolutely continuous and discrete components of the underlying random measure, and applying a result for Poisson processes known as the Palm formula \citep[e.g.][]{daley1988introduction,james2002poisson,CarTehMur2013a}, together with combinatorics. This is a very general approach, which can also be employed to derive the PMF of the Indian buffet process random binary matrix using  the beta-Bernoulli  process.

\subsection{The gamma-Poisson or negative binomial process}\label{sec:NBP0}
Let $\Nmat_J \sim\mbox{NBPM}(\gamma_0,c)$ denote a gamma-Poisson or negative binomial process (NBP) random count matrix, parameterized by a mass parameter $\gamma_0$ and a concentration parameter $c$.  This prior arises from marginalizing out the gamma process $G\sim\Gamma\mbox{P}(G_0,1/c)$ from $J$ conditionally independent Poisson process draws $X_j \mid G \sim \mbox{PP}(G)$, with the rows of $\Nmat_J$ corresponding to the $X_j$'s and the columns of $\Nmat_J$ corresponding to the atoms with at least one nonzero count. 

\subsubsection{Conditional likelihood}
As $\{X_j\}_{1,J}$ are i.i.d. given $G$, they are exchangeable according to de Fennetti's theorem. With a draw from the gamma-Poisson process expressed as $X_j=\sum_{k=1}^\infty n_{jk}\delta_{\omega_k},~n_{jk}\sim\mbox{Pois}(r_k)$, where $r_k=G(\omega_k)$ is the weight of the atom $\omega_k$ of the gamma process $G\sim\Gamma\mbox{P}(G_0,1/c)$, 
we may write the likelihood of $\{X_j\}_{1,J}$, given $G$, as
\begin{align}
p(\{X_j\}_{1,J} \mid G)
= \prod_{k=1}^\infty\frac{r_{k}^{n_{\cdotv k}}}{ \prod_{j=1}^J n_{jk}!}e^{-Jr_k}=  \Bigg\{ \prod_{k : n_{\cdotv k} > 0} \frac{r_{k}^{n_{\cdotv k}}}{ \prod_{j=1}^J n_{jk}!}e^{-Jr_k} \Bigg\} \cdot
\Bigg\{ \prod_{k : n_{\cdotv k} = 0} e^{-Jr_k} \Bigg\}\, ,
 \notag
\end{align}
where $n_{\cdotv k}=\sum_{j=1}^J n_{jk}$.  Let $\mathcal{D}_J=\{\omega_k\}_{k:n_{\cdotv k}>0}$ denote the set of all observed atoms with nonzero counts, and let $K_J = |\mathcal{D}_J|$.  Our goal is to marginalize out the random measure $G$ to obtain the unconditional PMF of the random count matrix $\Nmat_J\in\mathbb{Z}^{J\times K_J}$, where $\mathbb{Z}=\{0,1,\ldots\}$, and to show that this ``feature count'' matrix is 
row-column exchangeable.  The rows correspond to the $X_j$'s, and the $K_J$ columns represent those atoms in $\Omega$ with at least one nonzero count across the $X_j$'s. Representing the infinite dimensional $X_j$'s as a finite random matrix brings interesting  combinatorial questions that need to be carefully addressed. 

Fix an arbitrary labeling of the indices of the atoms in $\mathcal{D}_J$ from $1$ to $K_J$.  We now appeal to the definition of a gamma process and rewrite the conditional likelihood of $\{X_j\}_{1,J}$~as 
\begin{align}\label{eq:NBP_Like}
p(\{X_j\}_{1,J} \mid G) 
&= e^{-JG(\Omega\backslash\mathcal{D}_J) }\prod_{k=1}^{K_J}\frac{r_{k}^{n_{\cdotv k}}e^{-Jr_k}}{ \prod_{j=1}^J n_{jk}!}\, ,
\end{align}
where $G(\Omega\backslash\mathcal{D}_J):=\sum_{k:n_k=0}r_k$ is the total mass of the rest of the (absolutely continuous) space. 
The idea is to first marginalize out $G$ from (\ref{eq:NBP_Like}) to obtain the marginal distribution $p(\{X_j\}_{1,J} \mid \gamma_0, c)$, whose derivation using the Palm formula is provided in the Supplementary Material, and then use combinatorial argument to find the marginal distribution of the random count matrix $\Nmat_J$ organized from $\{X_j\}_{1,J}$. 

\subsubsection{Marginal distribution and combinatorial analysis}
One of our main results is that the PMF of $\Nmat_J \sim\mbox{NBPM}(\gamma_0,c)$, with $J$ rows and a random $K_J$ number of columns, is
\begin{align}\label{eq:NBP_PMF}
f(\Nmat_J \mid \gamma_0,c) &=\frac{p(\{X_j\}_{1,J} \mid \gamma_0, c)}{K_J!} 
= \frac{{\gamma_0^{K_J} \exp\left[-\gamma_0\ln(\frac{J+c}{c}) \right] }}{K_J!} \prod_{k=1}^{K_J} \frac{ \frac{\Gamma(n_{\cdotv k})}{(J+c)^{n_{\cdotv k}}}}{\prod_{j=1}^J n_{jk}!} \, ,
\end{align}
where the  unordered column vectors $\{\nv_{:k}\}_{1,K_J}$ of the count matrix $\Nmat_J$ represent a draw from the underlying stochastic process,
and the normalization constant of $1/K_J!$ arises from the fact that the mapping from a realization of $\{X_j\}_{1,J}$ 
to $\Nmat_J$ is one-to-many, with $K_J!$ distinct column orderings.

By construction, the rows of a NBP random count matrix are exchangeable. Moreover, one may verify by direct calculation that a NBP random count matrix with PMF (\ref{eq:NBP_PMF}) can be generated column by column as i.i.d. count vectors:
\begin{align}\label{eq:NBProwwise}
&\nv_{:k}  \sim \mbox{Multinomial}(n_{\cdotv k}, 1/J,\ldots,1/J),\notag \\ 
&n_{\cdot k}\sim\mbox{Log}[{J}/{(J+c)}],\notag\\
& K_J\sim \mbox{Pois}\left\{\gamma_0\left[\ln(J+c)-\ln(c)\right]\right\}\, .
\end{align}
It is clear from (\ref{eq:NBProwwise}) that the columns of $\Nmat_J$ are  
independent multivariate count vectors, which all follow the same logarithmic-multinomial (mixture) distribution.   Thus the NBP random count matrix $\Nmat_J$ is row-column exchangeable \citep[see, e.g.][for a general treatment of row-column exchangeable matrices]{Hoover,aldous,orbanz2013bayesian}.

Now consider the row-wise sequential construction of the NBP random matrix, recalling that $\Nmat^+_{J+1}$ represents the ``new'' part of the matrix added by the new row. 
With the prior on $\Nmat_J\in\mathbb{Z}^{J\times K_J}$ well defined, one may construct $\Nmat_J$ in a sequential manner as
$$
f(\Nmat_J\mid \thetav) = f(\Nmat_1\mid \thetav)\frac{f(\Nmat_2\mid \thetav)}{f(\Nmat_1\mid \thetav)}\ldots \frac{f(\Nmat_J\mid \thetav)}{f(\Nmat_{J-1}\mid \thetav)} \, ,
$$
where $\thetav:=\{\gamma_0,c\}$  and  $p(\Nmat^{+}_{j+1}\! \mid\! \Nmat_j, \thetav):= f(\Nmat_{j+1}\!\mid \!\thetav)/f(\Nmat_{j}\!\mid\! \thetav)$ is the prediction rule to add the new part brought by row $(j+1)$ into 
the matrix $\Nmat_j$.   Direct calculations using (\ref{eq:NBProwwise})  
yield the following form for this prediction rule,  
expressed in terms of familiar PMFs: 
\begin{align}
\label{eq:NBP_pre}
 p(\Nmat^{+}_{J+1} \mid \Nmat_J,\thetav) &= \frac{K_J! K^+_{J+1}!}{K_{J+1}!} 
 \prod_{k=1}^{K_J}\mbox{NB}
\left(n_{(J+1)k};n_{\cdotv k},\frac{1}{J+c+1}\right)\notag\\
&\times \prod_{k=K_J+1}^{K_{J+1}} \mbox{Log}\left(n_{(J+1)k};\frac{1}{J+c+1}\right)\notag\\
&\times \mbox{Pois}\left\{K^+_{J+1}; \gamma_0\left[\ln(J+c+1)-\ln(J+c)\right]\right\}.
\end{align}
This formula says that to add a new row 
to $\Nmat_J\in\mathbb{Z}^{J \times K_J}$, 
we first draw count $\mbox{NB}[n_{\cdotv k},1/(J+c+1)]$ at each existing column.  We then draw $K^+_{J+1}$ new columns as
$K^+_{J+1}\sim\mbox{Pois}\{ \gamma_0[\ln(J+c+1)-\ln(J+c)]\}$.  Finally, each entry in the new columns has a $\mbox{Log}[{1}/{(J+c+1)}]$ distributed random count; crucially, new columns brought by the new row must have positive counts. 

The normalizing constant $(K_J! \ K^+_{J+1}!) / K_{J+1}! $ in (\ref{eq:NBP_pre}) plays a key role in our combinatorial analysis, and will appear again in both the gamma- and  beta- negative binomial processes.  It emerges directly from the calculations, 
and can also be interpreted in the following way.  After drawing $K_{J+1}^+$ new columns, we must insert them into the original $K_J$ columns while keeping the relative orders of both the original and new columns unchanged.   This is a one-to-many mapping, with the number of such order-preserving insertions given by the binomial coefficient.  For example, if the original $\Nmat_J$ has  two columns and the new row $J+1$ introduces two more columns, then we construct $\Nmat_{J+1}$ by rearranging the two old columns 1 and 2 and the two new columns iii and iv in one of $\binom{4}{2}=6$ possible ways:  (1 2 iii iv), (1 iii 2 iv), (iii 1 2 iv), (1 iii iv 2), (iii 1 iv 2), and (iii iv 1 2), where  (1 2 iii iv) represents the construction appending the new columns to the right of the original matrix.

It is instructive to compare (\ref{eq:NBProwwise}), which generates a NBP random matrix by drawing all its columns at once,  with (\ref{eq:NBP_pre}), which generates an identically distributed random matrix one row at a time. 
The matrix generated with (\ref{eq:NBProwwise}) has i.i.d. columns. The matrix generated with  (\ref{eq:NBP_pre}) adds $K^+_{J+1}$ new columns when it adds the $(J+1)$th row, and if the newly added columns are inserted into random locations among original columns with their relative orders preserved, then we arrive at an identically distributed column-i.i.d. random count matrix. If the newly added columns are inserted in a particular way,  then the distribution of the generated random matrix would be different up to a multinomial coefficient.
For example,  if we generate row vectors $\nv_{j}$ from $j=1$ to $j=J$ and each time we append the new columns to the right of the original matrix, 
then this ordered matrix $\widetilde{\Nmat}_J$ 
will appear with probability 
\begin{align}\label{eq:Seq_NBPMatrix}
f(\widetilde{\Nmat}_J\mid \thetav)& =
f(\Nmat_1\mid \thetav)\prod_{j=1}^{J-1} p(\Nmat^{+}_{j+1} \mid \Nmat_j,\thetav)\frac{K_{j+1}!}{K_j! K^+_{j+1}!}
=\binom{K_J}{K^+_1,\ldots,K^+_J} f(\Nmat_J\mid \thetav).
\end{align}
Shown in the first row of Figure \ref{fig:NBPs_Matrix_Draw} are three NBP random count matrices simulated in this manner. 
We note that the gamma-Poisson process is related to the model of \citet{lo1982bayesian}, as well as the model of \citet{InfGaP}, which can be considered as a special case of the NBP with the concentration parameter $c$ fixed at one.

 \subsubsection{Inference for parameters}\label{sec:NBPinfer}
 
 Although the marginal likelihood alone is not amenable to posterior analysis, 
the NBP parameters can be conveniently  inferred  using both the conditional and marginal likelihoods.
 To complete the model, we let  $\gamma_0\sim \mbox{Gamma}(e_0, {1}/{f_0})$ and $c\sim\mbox{Gamma}(c_0,1/d_0)$. 
With (\ref{eq:NBP_Like}),  (\ref{eq:NBP_PMF}) and $G(\Omega):=G(\Omega\backslash\mathcal{D}_J) +  \sum_{k=1}^{K_J} r_k$, we sample the parameters in closed form as 
\begin{align}\label{eq:NBP_sampling}
&(\gamma_0 \mid -)\sim\mbox{Gamma}\bigg(e_0+K_J, \frac{1}{f_0-\ln(\frac{c}{c+J})}\bigg),\notag\\
&(r_k \mid -)\sim\mbox{Gamma}\big(n_{\cdotv k}, {1}/{(c+J)}\big),\notag\\
&\{G(\Omega\backslash\mathcal{D}_J) \mid -\} \sim \mbox{Gamma}\big(\gamma_0,{1}/{(c+J)}\big),\notag\\
&(c \mid -)\sim\mbox{Gamma}\big(c_0+\gamma_0,1/[d_0+G(\Omega)]\big) \, .
\end{align}
Similar strategies will be used to infer the parameters of the other two stochastic processes. 
Having closed-form update equations for parameter inference via Gibbs sampling 
is a unique feature shared by all the nonparametric Bayesian priors proposed in this paper. 

\subsection{The gamma-negative binomial process}\label{sec:GNBP0}
Let   $
  \Nmat_J\sim\mbox{GNBPM}(\gamma_0,c,p_1,\ldots,p_J)\notag
  $ denote a gamma-negative binomial process (GNBP) random count matrix, parameterized by  
a mass parameter $\gamma_0$, a concentration parameter $c$, and $J$ row-specific probability parameters $\{p_j\}_{1,J}$. This random count matrix is  the direct outcome of marginalizing out the gamma process $G \sim \Gamma \mbox{P}(G_0, 1/c)$, with data augmentation, from $J$ conditionally independent negative binomial process draws $X_j \mid G\sim\mbox{NBP}(G,p_j)$, which are defined such that $X_j(A) \sim\mbox{NB}\left(G(A),p_j\right)$ for each $A\subset\Omega$.

As directly marginalizing out the gamma process under negative binomial sampling is difficult, our construction is based on the compound-Poisson representation of the negative binomial, described in Section \ref{sec:prelims}.  Specifically, consider the joint distribution of $\Nmat_J$ and a latent count matrix $\Lmat_J$, whose dimension and locations of nonzero counts are the same as those of $\Nmat_J$.  These two matrices parallel the scalar $n$ and $l$ given in the joint PMF of the Poisson-logarithmic distribution (\ref{eqn:NBcompoundPoisson}).   This joint distribution is defined as
\beq\label{eq:marginal_gamma0}
f(\Nmat_J,\Lmat_J \mid \thetav) =
 \frac{\gamma_0^{K_J} \exp\left[-\gamma_0\ln(\frac{c+q_{\cdotv}}{c})\right]}{K_J!}\prod_{k=1}^{K_J} \frac{\Gamma( l_{\cdotv k})}{(c+q_{\cdotv} )^{ l_{\cdotv k}}} \left( \prod_{j=1}^J\frac{ |s(n_{jk},l_{jk})|p_j^{n_{jk}}}{n_{jk}!}\right)\, ,
\eeq 
where $\thetav:=\{\gamma_0,c,p_1,\ldots,p_J\}$, $q_j:=-\ln(1-p_j)$ and $q_{\cdotv} := \sum_{j=1}^J q_j$. The detailed derivation is in the Supplementary Material.

Similar to the analysis in Section \ref{sec:NBP0} for the NBP, we show in the Supplementary Material that the GNBP random count matrix can be constructed  by either drawing its i.i.d. columns  at once or  adding one row at a time, and it has closed-form Gibbs sampling update equations for model parameters. Different from 
the NBP random count matrix that is row-column exchangeable, the GNBP random count matrix no longer maintains row exchangeability if its row-wise probability parameters $p_j$ are set differently for different rows.

Shown in the second row of Figure \ref{fig:NBPs_Matrix_Draw} are three sequentially constructed GNBP random count matrices, with the new columns introduced by each row appended to the right of the matrix. 
Similar to the combinatorial arguments that lead to (\ref{eq:Seq_NBPMatrix}), this particularly structured matrix and its auxiliary matrix appear with probability $\binom{K_J}{K^+_1,\ldots,K^+_J} f(\Nmat_J,\Lmat_J\!\mid\! \thetav)$.


\subsection{The beta-negative binomial process}\label{sec:BNBP0}
Let 
$
\Nmat_J \sim\mbox{BNBPM}(\gamma_0,c,r_1,\ldots,r_J) \notag
$
denote a beta-negative binomial process (BNBP) random count matrix, parameterized by a mass parameter $\gamma_0$, a concentration parameter $c$, and $J$ row-specific dispersion parameters $\{r_j\}_{1,J}$, whose PMF 
is defined as 
\begin{align}\label{eq:BNBP_Matrix}
&f(\Nmat_J\mid \thetav)
 =   
 \frac{\gamma_0^{K_J}\exp\left\{-\gamma_0\left[\psi(c+r_{\cdotv})-\psi(c)\right]\right\}}{K_J!}
\prod_{k=1}^{K_J}  \frac{\Gamma(n_{\cdotv k})\Gamma(c+r_{\cdotv})}{\Gamma(c+n_{\cdotv k}+r_{\cdotv})}\prod_{j=1}^J  \frac{\Gamma(n_{jk}+r_j)}{n_{jk}!\Gamma(r_j)}\, ,
\end{align}
where $\thetav:=\{ \gamma_0,c,r_1,\ldots,r_J\}$.
The PMF is the direct outcome of marginalizing out the beta process $B \sim \mbox{BP}(c,B_0)$ from $J$ conditionally independent negative binomial process draws $X_j \!\mid\! B \sim \mbox{NBP}(r_j,B)$, which are defined such that $X_j(A) = \sum_{k:\omega_k\in A} n_{jk},~n_{jk} \sim\mbox{NB}(r_j,p_k)$ for each $A\subset\Omega$, where $p_k=B(\omega_k)$ is the weight of atom $k$. The detailed derivation is provided in the Supplementary Material. 

Similar to the analysis in Section \ref{sec:NBP0} for the NBP, we show in the Supplementary Material that the BNBP random count matrix can be constructed  by either drawing its i.i.d. columns  at once or  adding one row at a time using an ``ice cream'' buffet process, and it has closed-form Gibbs sampling update equations for all model parameters except for the concentration parameter~$c$. The BNBP random count matrix no longer maintains row exchangeability if its row-wise dispersion parameters $r_j$ are set differently for different rows.

Shown in the last row of Figure \ref{fig:NBPs_Matrix_Draw} are three sequentially constructed BNBP random count matrices, with the new columns introduced by each row appended to the right of the matrix. 
 Similar to the combinatorial arguments that lead to (\ref{eq:Seq_NBPMatrix}), this particularly structured matrix appears with probability $\binom{K_J}{K^+_1,\ldots,K^+_J} f(\Nmat_J\!\mid\!\thetav)$.

\subsection{The predictive distribution of a new row count vector}\label{sec:predict_vector}
It is critical to note that the prediction rule $p(\Nmat^{+}_{J+1} \!\mid\! \Nmat_J,\thetav)$ of the NBP shown in (\ref{eq:NBP_pre}) is for sequentially constructing a column-i.i.d. random count matrix, but it is not the predictive distribution for a new row count vector. 
 The $1\times K_J$ submatrix of $\Nmat^+_{J+1}$ orders its column in the same way as $\Nmat_J$ does, and the  $(J+1) \times K^+_{J+1}$ submatrix of $\Nmat^+_{J+1}$ also maintains a certain order of its columns; 
however, the indexing of these $K^+_{J+1}$ columns are in fact arbitrarily chosen from $K^+_{J+1}!$ possible permutations. 
Therefore, the predictive distribution of a row vector $\nv_{J+1}$ that brings $K^+_{J+1}$ new columns shall be
\begin{align}\label{eq:NBP_pre1}
p(\nv_{J+1} \mid \Nmat_{J},\thetav) &= 
 \frac{ p(\Nmat^{+}_{J+1} \mid \Nmat_J,\thetav) }{K^+_{J+1}!}
  \\ 
 &= {\frac{K_{J}!}{K_{J+1}! }  }\frac{\frac{K_{J+1}!}{K_J! K^+_{J+1}!} f(\Nmat_{J+1}\mid \thetav) }{f( \Nmat_J\mid \thetav)}\, .\label{eq:NBP_pre1_1}
\end{align}
The normalizing constant $1 / {K^+_{J+1}!}$ in (\ref{eq:NBP_pre1}) arises because a realization of $\Nmat^{+}_{J+1}$ 
to $\nv_{J+1}$ is one-to-many, with $K^+_{J+1}!$ distinct orderings of these new columns brought by the $(J+1)$th row.  Our experimental results show that omitting this normalizing term may significantly deteriorate the out-of-sample prediction performance. 

An equivalent representation in (\ref{eq:NBP_pre1_1}) shows that one may first consider the distribution of a matrix constructed by appending the new columns brought by $\nv_{J+1}$ to the right of $\Nmat_J$, which is $\frac{K_{J+1}!}{K_J! K^+_{J+1}!} f(\Nmat_{J+1}\!\mid\! \thetav) $, and then apply the Bayes' rule to derive the conditional distribution of this particularly ordered $\nv_{J+1}$ given  $\Nmat_J$. 
The normalizing constant ${K_J! } / {K_{J+1}!}$ in  (\ref{eq:NBP_pre1_1}) 
can be interpreted in the following way.   We need to insert the $K^+_{J+1}$ new columns one by one into the original matrix. The first, second, $\ldots$, and last new columns can choose from $K_J+1$, $K_J+2$, $\ldots$, and $K_J+K^+_{J+1}$ possible locations, respectively, thus there are $\prod_{i=1}^{K^+_{J+1}}(K_J+i)! = {K_{J+1}!} / {K_{J}!}$ ways to insert the   $K^+_{J+1}$ new columns into the original ordered $K_{J}$ columns, which is again a one-to-many mapping.
The same combinatorial analysis applies to both the GNBP and BNBP. For the GNBP, to compute the predictive likelihood of $\nv_{J+1}$, one will need to take extra care as the computation involves $\Lmat_J$, an auxiliary random count matrix that is not directly observable. 
In Section \ref{sec:experiments}, we will discuss in detail  how to compute the predictive likelihood via Monte Carlo integration.

\subsection{Comparison}

In the Supplementary Material, we provide further details on the construction of random count matrices from the negative binomial process, as well as those derived from 
both the gamma-negative binomial process (GNBP) and  beta-negative binomial process (BNBP).  While the PMFs for all three proposed nonparametric priors are complicated, 
 their relationship and differences become evident once we  show that they all govern random count matrices with a Poisson-distributed number of  i.i.d.~columns. Table~\ref{tab:NBP} shows the differences among the three priors' row-wise sequential construction, and the following list shows the variance-mean relationship for each prior for the counts at existing columns.  Together, these provide additional insights on how the priors differ from each other. 
\begin{eqnarray}
\mbox{NBP:} \quad  \mbox{Var}[n_{(J+1)k}] &=&  \E[n_{(J+1)k}]+\frac{\E^2[n_{(J+1)k}]}{n_{\cdotv k}} \label{eqn:varmeanNBP} \\
\nonumber \\
\mbox{GNBP:} \quad  \mbox{Var}[n_{(J+1)k}] &=&  \frac{\E[n_{(J+1)k}]}{1-p_{J+1}} + \frac{ \E^2[n_{(J+1)k}]}{l_{\cdotv k}} \label{eqn:varmeanGNBP} \\
\nonumber \\
\mbox{BNBP:} \quad  \mbox{Var}[n_{(J+1)k}] &=&  \frac{\E[n_{(J+1)k}]}{\frac{c+r_{\cdotv} }{n_{\cdotv k}+c+r_{\cdotv} -1}} + \frac{ \E^2[n_{(J+1)k}]}{\frac{n_{\cdotv k}(c+r_{\cdotv} -2)}{ n_{\cdotv k}+c+r_{\cdotv} -1}} \label{eqn:varmeanBNBP}
\end{eqnarray}

\begin{table}[t]
\begin{footnotesize}
\caption{Comparison of the prediction rules of the NBP, GNBP, and BNBP random count matrices.}\label{tab:NBP}
\vspace{-4mm}
\begin{center}
\begin{tabular}{r  p{14pc} p{10pc} p{9pc} }
\toprule
Model &Number of new columns $K^+_{J+1}$ & Counts in existing columns & Counts in new columns \\
\midrule
NBP& $\mbox{Pois}\left\{\gamma_0[\ln(J+c+1)-\ln(J+c)]\right\}$ &  $ \mbox{NB}\left[n_{\cdotv k}, {1}/{(J+c+1)}\right] $ &  $ \mbox{Log}\left[{1}/{(J+c+1)}\right] $ \\
GNBP& $\mbox{Pois}\left\{\gamma_0 \left[\ln(c+q_{\cdotv}+q_{J+1}) - \ln(c+q_{\cdotv}) \right]\right\}$ & $\mbox{GNB}\left(l_{\cdotv k},c+q_{\cdotv} , p_{J+1} \right)$& $\mbox{LogLog}\left(c+q_{\cdotv} , p_{J+1} \right)$  \\ 
BNBP& $\mbox{Pois}\left\{\gamma_0\left[\psi(c+r_{\cdotv}+r_{J+1})-\psi(c+r_{\cdotv})\right]\right\}$ & $\mbox{BNB}(r_{J+1},n_{\cdotv k}, c+r_{\cdotv})$& $\mbox{Digam}(r_{J+1}, c+r_{\cdotv})$  \\
\bottomrule
\end{tabular}
\end{center}
\end{footnotesize}
\end{table}%


\begin{figure}[!tb]
\begin{center}
\includegraphics[width=.88\columnwidth]{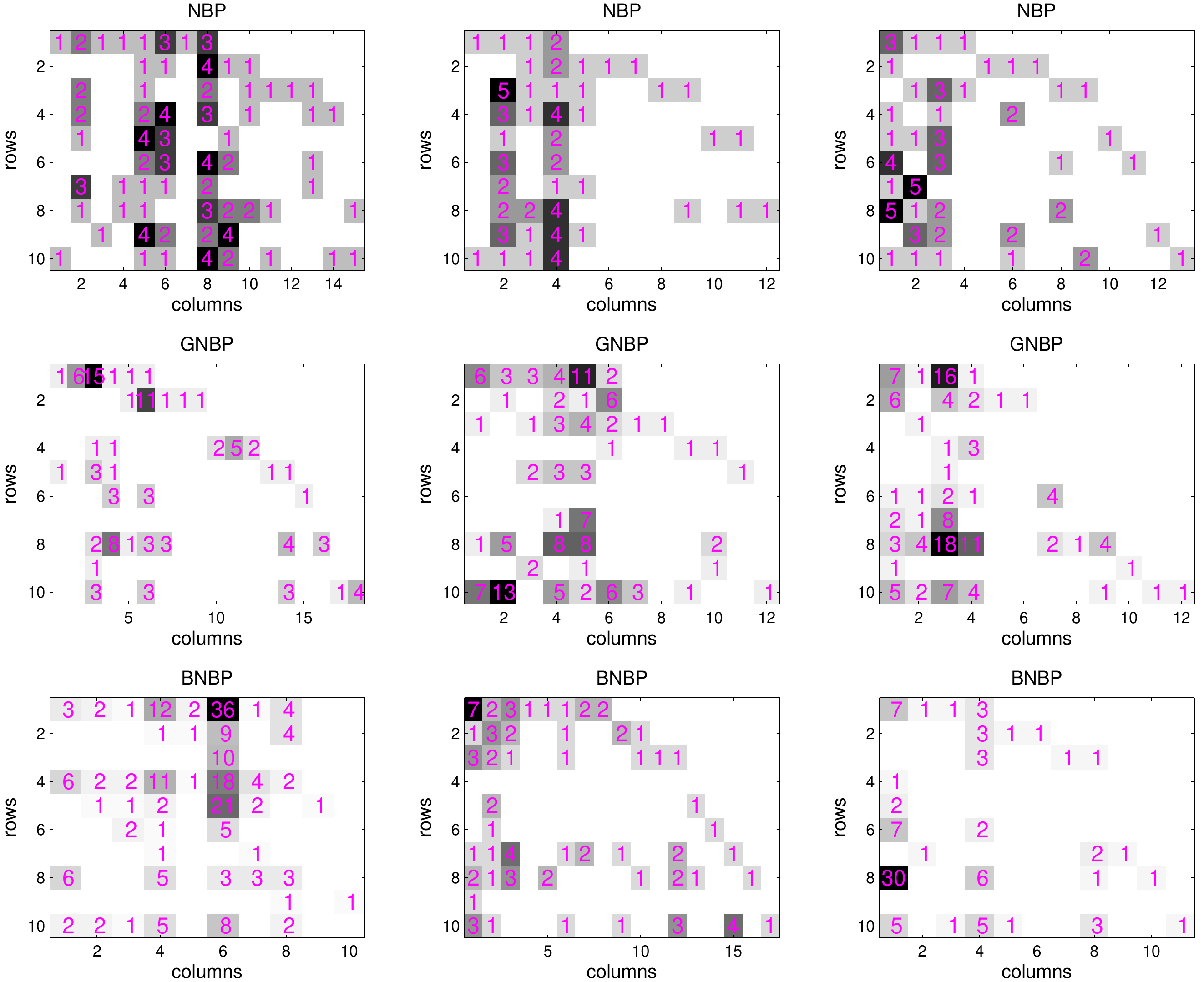}
\end{center}
\vspace{-4mm}
\caption{ \label{fig:NBPs_Matrix_Draw}  
Sequentially constructed negative binomial process (NBP), gamma-negative binomial process (GNBP), and beta-negative binomial process  (BNBP) random count matrices (the blank cells indicate zero counts). 
The ten rows of each matrix are added one by one,
with  the new columns introduced by each row appended to the right of the matrix.
To make the expected total count of a random matrix as $100$ and the expected  number of columns approximately as $12$, 
the parameters are set as $\gamma_0=5$ and $c=0.5$ for the NBP, set as $c=1$, $\gamma_0=4.79$, and $\sum_{j} \frac{p_j}{1-p_j}=  20.88$ for the GNBP, and set as  $c=2$, $\gamma_0=4.31$, and $\sum_{j}  r_{j} = 23.20$ for the BNBP. The randomized row wise parameters $[p_1/(1-p_1),\ldots,p_J/(1-p_J)]^T$ and $(r_1,\ldots,r_J)^T$ are generated via   $ \mbox{Dir}(1,\ldots,1) \sum_{j} \frac{p_j}{1-p_j}$ and $\mbox{Dir}(1,\ldots,1)\sum_{j} r_j $, respectively.
}
\end{figure}

The NBP 
can be used to generate a row-column exchangeable random count matrix with a potentially unbounded number of columns. However, as shown in (\ref{eq:NBProwwise}),   to model the total count of a column $n_{\cdotv k}$, the NBP uses the logarithmic distribution, which   has only one free parameter, always has the mode at one, and monotonically decreases. In addition, each column sum $n_{\cdotv k}$ is assigned to the $J$ rows with a multinomial distribution that has a uniform probability vector $(1/J,\ldots,1/J)$. Furthermore, as shown in Table \ref{tab:NBP}, for out-of-sample prediction, it models counts at existing columns using $\mbox{NB}
\left[n_{(J+1)k};n_{\cdotv k},{1}/{(J+c+1)}\right]$, whose variance-mean relationship (\ref{eqn:varmeanNBP}) may be restrictive in modeling highly overdispersed counts.
Finally, the expected number of new columns brought by a row, equal to $\gamma_0\ln[1+{1}/{(J+c)}]$, monotonically decreases. 
These constraints limit the potential use of the NBP  model.  

Both the GNBP and BNBP relax these constraints in their own unique ways.  
 Examining the sequential construction of the GNBP 
  helps us understand the advantages of the GNBP over the NBP.
As shown in Table  \ref{tab:NBP}, to model the likelihood of a new row count vector, one may find that the GNBP employs the three-parameter GNB instead of the two-parameter negative binomial distribution to model the count at an existing column, and employs the two-parameter LogLog 
instead of the logarithmic distribution 
to model the count at a new column. As the GNB random variable $n_{(J+1)k}\sim\mbox{GNB}\left(l_{\cdotv k},c+q_{\cdotv} , p_{J+1} \right)$ can be generated as
$n_{(J+1)k}\sim\mbox{NB}(r_{(J+1)k},p_{J+1}),~r_{(J+1)k}\sim \mbox{Gamma}\left[l_{\cdotv k},1/(c+q_{\cdotv})  \right]$, using the laws of total expectation and total variance, we express $\mbox{Var}[n_{(J+1)k}]$ in terms of $\E[n_{(J+1)k}]$ in (\ref{eqn:varmeanGNBP}). 
Since $p_{J+1}<1$ and $l_{\cdotv k}\le n_{\cdotv k}$, the GNBP can model much more overdispersed counts than the NBP.
Moreover, the GNBP allows each row count vector to have its own probability parameter, allowing finer control on the expected number of new columns brought by a new row, which is $\gamma_0 \ln[1+{q_{J+1}}/{(c+q_{\cdotv})}]$.  The NBP random count matrix is row-column exchangeable, whereas the GNBP random count matrix is column exchangeable, but not row exchangeable if the row-wise probability parameters $p_j$ are fixed at different values.

As shown in Table  \ref{tab:NBP}, to model the likelihood of a new row count vector, one may find that the BNBP employs the three-parameter BNB instead of the two-parameter negative binomial distribution to model the count at an existing column, and employs the two-parameter digamma instead of the logarithmic distribution to model the count at a new column. Note that the BNB random variable $n_{(J+1)k}\sim\mbox{BNB}(r_{J+1},n_{\cdotv k}, c+r_{\cdotv})$ can be generated as
$n_{(J+1)k}\sim\mbox{NB}(r_{J+1},p_{(J+1)k}),~p_{(J+1)k}\sim \mbox{Beta}\left(n_{\cdotv k},c+r_{\cdotv}  \right)$, using the laws of total expectation and total variance, for $c+r_{\cdotv} >2$, 
we express $\mbox{Var}[n_{(J+1)k}]$ in terms of $\E[n_{(J+1)k}]$ in  (\ref{eqn:varmeanBNBP}). 
 As $\frac{c+r_{\cdotv} }{n_{\cdotv k}+c+r_{\cdotv} -1}\hspace{-.3mm}\le1$ and $\frac{n_{\cdotv k}(c+r_{\cdotv} -2)}{ n_{\cdotv k}+c+r_{\cdotv} -1}<n_{\cdotv k}$ for $c+r_{\cdotv} >2$,
the BNBP can also model much more overdispersed counts than the NBP.  Moreover, the BNBP allows each row count vector to have its own dispersion parameter, allowing finer control on the expected number of new columns brought by a row, which is $\gamma_0[\psi(c+r_{\cdotv}+r_{J+1})-\psi(c+r_{\cdotv})]$; the NBP random count matrix is row-column exchangeable, whereas the BNBP random count matrix is column exchangeable, but not row exchangeable if the row-wise dispersion parameters $r_j$ are different. 

The variance-mean relationships expressed by (\ref{eqn:varmeanNBP})-(\ref{eqn:varmeanBNBP}) show that the GNBP and BNBP can model much more overdispersed counts than the NBP.  This fact is borne 
out by the simulated random count matrices in Figure \ref{fig:NBPs_Matrix_Draw}, which provide some intuition for the practical differences among the models.  The parameters for the three priors have been chosen so that each random matrix has the same expected total count.  Yet the counts in the NBP random count matrices have small dynamic ranges, whereas the counts in both the GNBP and BNBP matrices can contain values that are significantly above the average.

 \subsection{Parameter inference}
An appealing feature of all three negative binomial process random count matrix priors is that their parameters can be inferred with closed-form Gibbs sampling update equations, by exploiting both the conditional and marginal distributions, together with the data augmentation and marginalization techniques unique  to the negative binomial distribution. 
Parameter inference for the NBP is provided in Section \ref{sec:NBPinfer}.  The details of parameter inference for both the GNBP and BNBP are provided  in the Supplementary Material.


%
\section{Negative Binomial Process Naive Bayes Classifiers}\label{sec:experiments} 

\subsection{Background}

Given a  random count matrix, finding the predictive distribution of a  row count vector, which  may bring additional columns,
involves interesting and challenging combinatory arguments that have been throughly addressed in this paper. With these combinatorial structures carefully analyzed, we are ready to construct a NBP, a GNBP, and a BNBP naive Bayes classifiers.  We do so as follows.  First, for each category, the training row count vectors are summarized as a random count matrix $\Nmat_J$, each column of  which must contain at least one nonzero count (i.e.~columns with all zeros are excluded).   Second, Gibbs sampling is used to infer the parameters $\thetav$ that generate $\Nmat_J$.  To represent the posterior of $\thetav$, $S$ MCMC samples $\{\thetav^{[s]}\}_{1,S}$ are collected.   For the GNBP, a posterior MCMC sample $\Lmat_J^{[s]}$ for the auxiliary random matrix is also collected when $\thetav^{[s]}$ is collected. 
Finally, to test a row count vector $\nv_{J+1}$, its predictive likelihood given $\Nmat_J$ 
is  calculated via Monte Carlo integration using 
\beq 
p(\nv_{J+1} \mid \Nmat_{J}) =\frac{1}{S} \sum_{s=1}^{S}\frac{ p(\Nmat^{+}_{J+1} \mid \Nmat_J, \thetav^{[s]}) }  { K^+_{J+1}!} \label{eq:prediction1}
\eeq 
for both the NBP and BNBP, and using \beq p(\nv_{J+1} \mid \Nmat_{J}) = \frac{1}{S}
  \sum_{s=1}^{S} \frac{ p(\Nmat^{+}_{J+1} \mid \Nmat_J, \Lmat_J^{[s]}, \thetav^{[s]}) } {K^+_{J+1}!}\label{eq:prediction2}\eeq for the GNBP. Although a larger $S$ shall lead to a more accurate calculation of the predictive likelihood, the computational complexity for testing is a linear function of $S$. It is therefore of practical importance  to find out how the value of $S$ impacts the performance of the proposed nonparametric Bayesian naive classifiers. Below we consider experiments on document categorization, for which we will show that $S=1$ performs essentially just as well as selecting a much larger $S$ in terms of the categorization accuracy.

  

\subsection{Experiment settings}
We consider the example of categorizing the 18,774 documents of the 20 newsgroups dataset\footnote{\href{http://qwone.com/~jason/20Newsgroups/}{http://qwone.com/$\sim$jason/20Newsgroups/}}, where each bag-of-words document is represented as a word count vector under a vocabulary of size $V= $ 61,188. We also consider  the TDT2 corpus\footnote{\href{http://www.cad.zju.edu.cn/home/dengcai/Data/TextData.html}{http://www.cad.zju.edu.cn/home/dengcai/Data/TextData.html}} ( NIST Topic Detection and Tracking corpus): with the documents appearing in two or more categories removed, this subset of TDT2 consists of  9,394 documents  from the largest 30 categories, with a vocabulary of size $V=$ 36,771;  this dataset was used to compare document clustering algorithms in \citet{CHH05}. We train all three negative binomial processes using 10\%, 20\%, $\ldots$, or 80\% of the documents in each newsgroup of the 20 newsgroups dataset, and in each category of the TDT2 corpus. We then test on the remaining documents. We report our results based on five random training/testing partitions.
 
 To make comparison to other commonly used text categorization algorithms, we also consider a default setting for the 20 newsgroups dataset: using the first 11,269 documents for training and the other 7,505 documents collected at later times for testing. For this setting, we reports our results based on five  independent runs with random initializations. This allows us to compare our performance to many other papers that have proposed text classification algorithms and benchmarked their methods using this same split of the 20 newsgroups dataset.
 
 For the $i$th newsgroup/category  with $J^{(i)}$ training documents,  we construct a document-term count matrix $\Nmat^{(i)}_{J^{(i)}}\in\mathbb{Z}^{J^{(i)}\times K_{J^{(i)}}}$, whose element $n^{(i)}_{jk}$ represents the number of times term $k$ appearing in document $j$. Since only the terms present in the training documents of the $i$th category are considered, the column indices of $\Nmat^{(i)}_{J^{(i)}}$ correspond to the terms that appear at least once in training. 
 We use $x^{(i)}$ to denote that $x$ is a parameter inferred from $\Nmat^{(i)}_{J^{(i)}}$. Note that the column indices of $\Nmat^{(i)}_{J^{(i)}}$  can be arbitrarily ordered, which affects neither training nor out-of-sample prediction as long as their corresponding features are recorded. 
 

We collect $S$ MCMC samples of model parameters and auxiliary variables to compute the predictive likelihood for a new row count vector. 
In this paper, we run $S$ independent Markov chains and collect the 2500th sample of each chain. Note that one may also consider collecting $S$ samples at a certain interval from a single Markov chain after the burn-in period.  
We consider non-informative hyper-parameters as $a_0=b_0=\ldots=f_0=0.001$. For the BNBP, we set $c_0=d_0=1$.  
The document-term training count matrix of the $i$th newsgroup is modeled as $\Nmat^{(i)}_{J^{(i)}}\sim\mbox{NBPM}(\gamma^{(i)}_0,c^{(i)})$, $\Nmat^{(i)}_{J^{(i)}}\sim\mbox{GNBPM}\big(\gamma^{(i)}_0,c^{(i)},p^{(i)}_1,\ldots,p^{(i)}_{J^{(i)}}\big)$, and $\Nmat^{(i)}_{J^{(i)}}\sim\mbox{BNBPM}\big(\gamma^{(i)}_0,c^{(i)},r^{(i)}_1,\ldots,r^{(i)}_{J^{(i)}}\big)$ under the three priors respectively. 

 Note that we are facing typical ``small $n$ and large $p$'' problems as the number of rows of a document-term count matrix is typically much smaller than the number of columns. For example, the first newsgroup of the 20 newsgroups dataset contains 798 documents with 12,665 unique words, which is summarized as a  $798\times 12665$ count matrix; and the 30th category of the TDT2 subset contains  52 documents with 2904 unique words, which is summarized as a  $52\times2904$ count matrix. 
 As the number of unique terms in a category might be significantly smaller than the vocabulary size of the whole corpus,  our approach for both training and  testing could  be much faster than the  approach that considers all the terms in the vocabulary of the corpus. In addition, our approach provides a principled, model-based way to handle terms that appear in a testing document but not in the training documents.  By contrast, many traditional approaches have to  discard these terms not present in training.

 \subsection{Training and posterior predictive checking}

 \begin{figure}[!tb]
\begin{center}
\includegraphics[width=0.98\columnwidth]{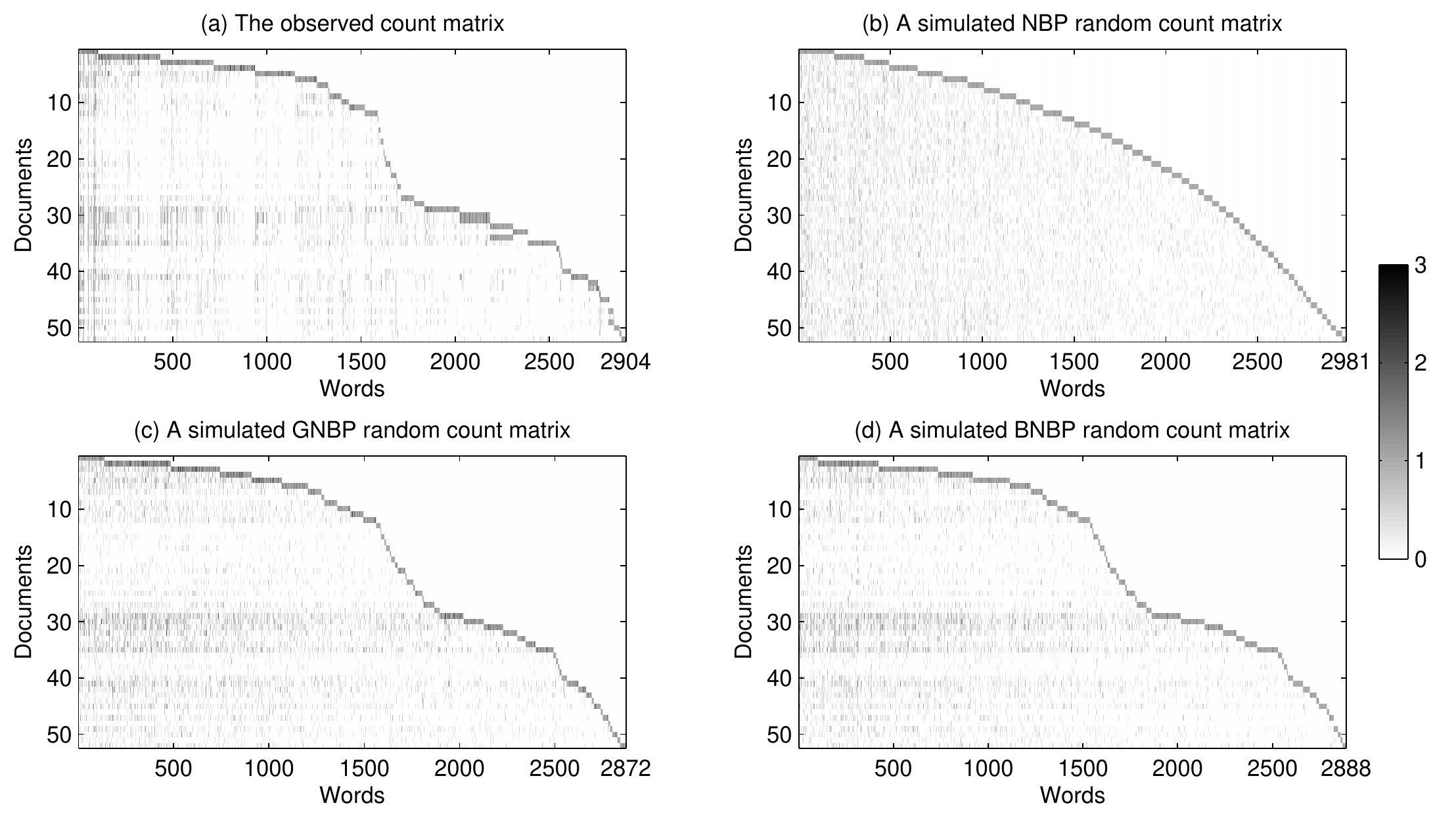}
\end{center}
\vspace{-3.5mm}
\caption{ \label{fig:generate}  
The parameters of the negative binomial processes are inferred using (a) the observed document-term count matrix. These parameters are used to simulate  (b) a NBP random count matrix, (c) a GNBP random count matrix, and (d) a BNBP random count matrix.  These matrices are visualized by arranging the new columns brought by each new row to the right of the original matrix. The counts larger than 3 are displayed as 3. 
}
\end{figure}

 We train the NBP, GNBP, and BNBP with the document-term word count matrix $\Nmat \in \mathbb{Z}^{52\times 2904}$ that summarizes all the 52 documents in the 30th category of the TDT2 subset. We then run 2500 MCMC iterations and collect the last 1500 samples to infer  the posterior means of the parameters in  $\Nmat \sim\mbox{NBPM}(\gamma_0,c)$, $\Nmat \sim\mbox{GNBPM}(\gamma_0,c,p_1,\ldots,p_{52})$, and $\Nmat \sim\mbox{BNBPM}(\gamma_0,c,r_1,\ldots,r_{52})$.  Using the corresponding parameters learned from the training count matrix,   we regenerate a NBP, a GBNP, and a BNBP random count matrix as an informal posterior predictive check on the model.  The observed count matrix is  shown in Figure \ref{fig:generate} (a), and the three simulated random count matrices are shown in Figure \ref{fig:generate} (b)-(d). These matrices are displayed by arranging  the new columns brought by a new row to the right of the original matrix. 
 
It is clear that the NBP is restrictive, in that the generated random matrix looks the least similar to the observed count matrix. 
This is unsurprising, as the NBP has a limited ability to model highly overdispersed counts, does not model row-heterogeneity, and can barely adjust the number of new columns brought by a row.   On the other hand, both the generated GNBP and BNBP random count matrices resemble the original count matrix much more closely.  This is expected, since both priors use heavy-tailed count distributions to model highly overdispersed counts,  and have row-wise probability or dispersion parameters to model row-heterogeneity and to control the number of new columns brought by each row. Note that the observed matrix has 2904 columns, but each of the generated random count matrices has a different (random) number of columns. This is because there are one-to-one correspondences between their row indices, but not their column indices.


\subsection{Out-of-sample prediction and categorization for count vectors}

For out-of-sample prediction on a new row vector, we first compute that vector's likelihood under different categories' training count matrices.  We then use these likelihoods in a naive-Bayes classifier to categorize the new vector. For example, for testing row count vector $\nv_{j'}$ under category $i$, we will first match the column indices (features) of this row count vector to those of the training count matrix $\Nmat^{(i)}_{J^{(i)}}$; each feature that belongs to one of the $K^{(i)}_{J^{(i)}}$ features of  $\Nmat^{(i)}_{J^{(i)}}$ but not present in $\nv_{j'}$ will be assigned a zero count;  and the ${K^+_{j'}}^{(i)}$ features  that are present in vector $j'$ but not in $\Nmat^{(i)}_{J^{(i)}}$ will be treated as new features brought by vector $j'$ to 
to $\Nmat^{(i)}_{J^{(i)}}$.  For the
 the GNBP, we first find an estimate of $p^{(i)}_{j'}$  as $
p^{(i)}_{j'} = {(a_0+n^{(i)}_{j'\cdotv})}/{[a_0+b_0+n^{(i)}_{j'\cdotv}+G^{(i)}(\Omega)]}$.  
For the BNBP,  we first find an expectation-maximization  estimate of $r_{j'}$ by running the updates
\begin{align}
&l^{(i)}_{j'k} = r^{(i)}_{j'}\big[\psi(r^{(i)}_{j'}+n^{(i)}_{j'k})-\psi(r^{(i)}_{j'})\big],\notag\\
&r^{(i)}_{j'} = \frac{a_0-1+ l^{(i)}_{j'\cdotv}}{b_0+ p^{(i)}_* -\sum_{k=1}^{K_{J^{(i)}}} \ln(1-p^{(i)}_k)}\notag
\end{align}
iteratively for 20 iterations, where for a testing row vector with all zeros, we let  $l^{(i)}_{j'\cdotv}=1$. 
Given the column sums of  $\Nmat^{(i)}$ and the inferred model parameters (together with auxiliary variables for the GBNB), the predictive likelihoods of a new row count vector  are calculated using  (\ref{eq:prediction1}) for both the NBP and BNBP and with (\ref{eq:prediction2}) for the GNBP.

%
%
%
%

Note that when the predictive distributions are used to calculate the likelihoods, the models are not constrained under a predetermined vocabulary. But if we are given a vocabulary of size $V$ that includes all the important terms, exploiting that information might further improve the performance. 
Thus  
to test 
document $j'$, 
we also consider using
\beq\label{eq:TestNBP}
p(\nv_{j'} \mid \Nmat^{(i)}_{J^{(i)}},\thetav^{(i)}) = \prod_{v=1}^{V}\mbox{NB}\left[n_{j'v};n^{(i)}_{\cdotv v}+\gamma_0^{(i)}/V, 1/(J^{(i)}+c^{(i)}+1)\right]
\eeq
as the likelihood for the NBP, 
using
\beq\label{eq:TestGNBP}
p(\nv_{j'}  \mid \Nmat_{J^{(i)}}^{(i)},\Lmat_J^{(i)},\thetav^{(i)}) = \prod_{v=1}^{V}\mbox{GNB}\left(n_{j'v};l^{(i)}_{\cdotv v}+\gamma^{(i)}_0/V, c^{(i)}+q^{(i)}_{\cdotv},p^{(i)}_{j'}\right)
\eeq
as the likelihood for the GNBP, 
and using
\beq\label{eq:TestBNBP}
p(\nv_{j'} \mid \Nmat_{J^{(i)}}^{(i)},\thetav^{(i)}) = \prod_{v=1}^{V}\mbox{BNB}\left(n_{j'v};r^{(i)}_{j'},n^{(i)}_{\cdotv v}+\gamma^{(i)}_0/V, c^{(i)}+r^{(i)}_{\cdotv}\right)
\eeq
as the likelihood for the BNBP. 
 Note that for this testing procedure we also  compute $p(\nv_{j'}  \mid \Nmat_{J}^{(i)})$ using Monte Carlo integration based on $S$ posterior MCMC samples.
In contrast to its truly nonparametric Bayesian counterpart with an infinite vocabulary, this testing procedure is expected to have higher computational complexity, but may produce  better out-of-sample prediction if the predetermined finite vocabulary fits the testing documents well.   Below we show the results produced by both testing procedures.  

For comparison,  we consider the multinomial naive Bayes classifier with Laplace smoothing \citep{mccallum1998comparison,manning2008introduction}, where a test document $j'$ has the likelihood under newsgroup $i$ as
\beq
\prod_{v=1}^{V} \left(\frac{n^{(i)}_{\cdot v}+1}{\sum_{v=1}^V(n^{(i)}_{\cdot v}+1)}\right)^{n_{j'v}}.
\eeq
The results of some other commonly used text classification algorithms will also be included as benchmarks.  Note that all these classifiers require the same predefined finite vocabulary for both training and testing.  Thus any new terms in a testing document that are not listed in that vocabulary must be discarded.

\begin{figure}[!tb]
\begin{center}
\includegraphics[width=0.98\columnwidth]{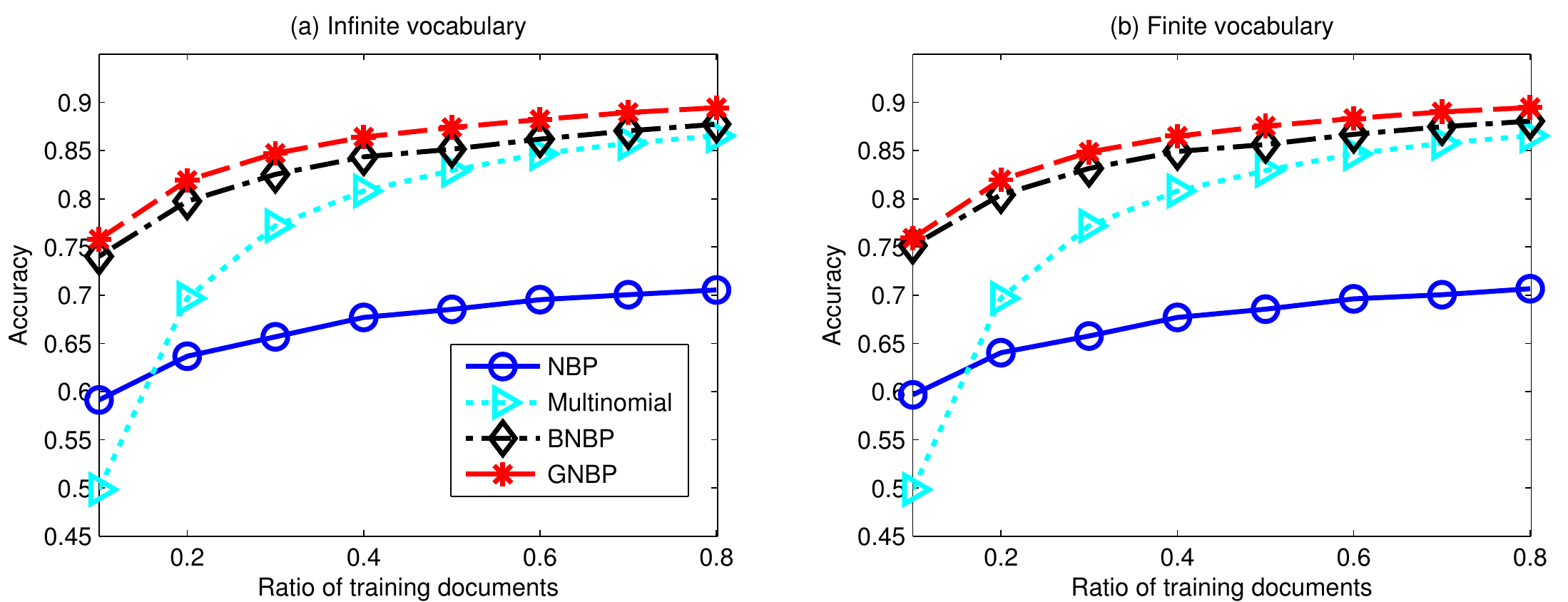}
\end{center}
\vspace{-3.5mm}
\caption{ \label{fig:20newsgroup}  
Document categorization results  on the 20 Newsgroup dataset with (a) an unconstrained  vocabulary that can grow to infinite, and (b) a predetermined finite vocabulary of size $V=$ 61,188, using the negative binomial process (NBP), gamma-negative binomial process (GNBP), and beta-negative binomial process (BNBP). The results of the multinomial naive Bayes classifier using Laplace smoothing are included for comparison. 
}

\begin{center}
\includegraphics[width=0.98\columnwidth]{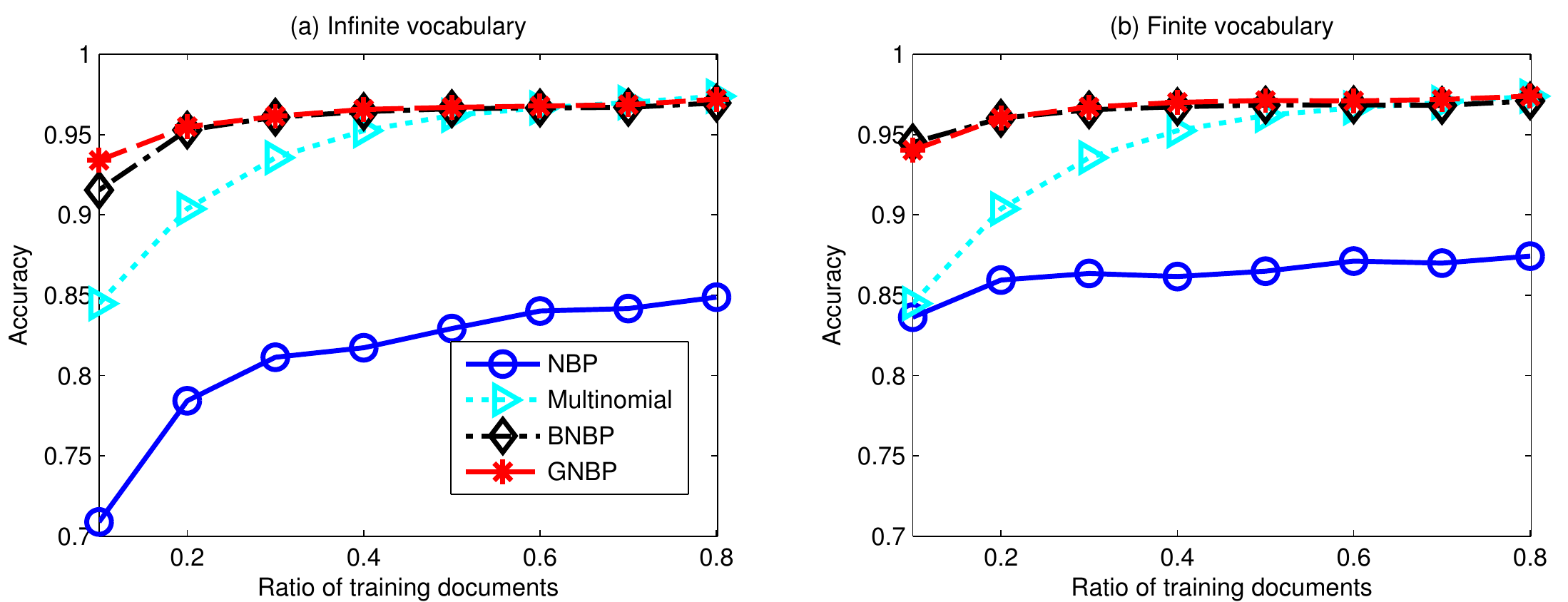}
\end{center}
\vspace{-3.5mm}
\caption{ \label{fig:TDT2}  
Analogous plots to Figures \ref{fig:20newsgroup} (a) and (b) for the TDT2 dataset. The predetermined finite vocabulary has the size of $V=$ 36,771.
}
\end{figure}

\subsection{Example results}

We first consider  choosing $S=10$  in (\ref{eq:prediction1})  and (\ref{eq:prediction2})  to compute the predictive likelihood $p(\nv_{j'}  \!\mid\! \Nmat_{J}^{(i)})$ for test document $j'$. 
Assuming a uniform prior for all the $C$ categories, we assign 
 document $j'$ to category $i$ with probability 
 \beq
\frac{p(\nv_{j'}  \mid \Nmat_{J}^{(i)})}{\sum_{i=1}^C p(\nv_{j'}  \mid \Nmat_{J}^{(i)})}
\eeq
and categorize document $j'$ to the category under which its word count vector $\nv_{j'}$ has the highest probability. 
As shown in Figures \ref{fig:20newsgroup} and \ref{fig:TDT2}, the NBP has the worst categorization accuracy.  Both the BNBP and GNBP clearly outperform the NBP and the multinomial naive-Bayes classifier with Laplace smoothing, especially when the number of training documents is small.  Both for fitting the training count matrix and making out-of-sample prediction, the NBP is the most restrictive, as it has only two free parameters $\gamma_0$ and $c$.   In addition to these two parameters, the GNBP (BNBP) has a probability (dispersion) parameter for each row count vector.
  Moreover, as both the GNB and BNB distributions are mixed negative-binomial distributions, they have heavier tails that  may help model the burstiness of words in documents 
  \citep{church1995poisson,madsen2005modeling,clinchant2008bnb}. 
  
  For the 20 newsgroups dataset, with the 7,505  documents collected at later times used for testing, our NBP, BNBP, and GNBP with an infinite vocabulary and $S=10$ achieve categorization accuracies of 61.9\%, 78.7\%, and 80.9\%, respectively.  With a finite vocabulary they achieve accuracies of 61.7\%, 79.1\%, and 80.9\%, respectively.  Despite the simplicity of the model, this performance meets or exceeds that of other competing methods, which we briefly describe.  The multinomial naive Bayes classifier with Laplace smoothing achieves an accuracy of 78.1\%.    \citet{lan2009supervised} consider a range of reweighted term-frequency features in a $k$-nearest neighbors ($k$NN) classifier.  Under an optimal choice of $k$ and set of features, they achievs an accuracy of 69.1\%.  The same authors report that a support vector machine (SVM) classifier achieves an accuracy of 80.8\%.  \citet{larochelle2012learning} use restricted Boltzmann machine for classification, with an optimized training strategy and cross-validated model parameters.  They report an accuracy of 76.2\% using binary features for the 5000 most frequent words.  The accuracy increases to 79.1\% when using binary features for the 25247 most frequent words, but the algorithm is too computationally intensive to include more word features.
  
 We also note that text categorization performance significantly deteriorates if one trains a multi-class classifier on the lower-dimensional  features extracted using unsupervised feature learning algorithms, such as latent Dirichlet allocation (LDA) \citep{LDA} or the deep Boltzmann machine   \citep{srivastava2013modeling}. As shown in  \citet{srivastava2013modeling}, even with tuned parameters, neither LDA nor deep Boltzmann machines combined with a multinomial logistic regression classifier  can achieve an accuracy above 70\% on this data set. It is also shown in \citet{zhu2012medlda} that LDA plus an SVM classifier fails to achieve an accuracy above 65\%.    The performance of LDA could be improved by using a supervised training strategy \citep{mcauliffe2008supervised}. However, as shown in \citet{zhu2012medlda}, the maximum-entropy discrimination LDA (MedLDA), a state-of-the-art supervised LDA algorithm, still does not achieve an accuracy above 80\%, despite the fact that the number of topics and model parameters are carefully tuned through cross validation and complex inference and heavy computations are employed to learn the latent features.  Both the BNBP and GNBP naive classifiers, while being tuning-free and fast and simple to train using the raw counts, compare favorably to the state-of-the-art text classification algorithms that often rely on heavy computation and carefully selected features and parameters.  
      
Note that for the proposed naive Bayes classifiers, a larger $S$ usually leads to a more accurate  computation of the predictive likelihood via Monte Carlo integration,  but may not necessarily lead to a clear gain in accuracy for document categorization. This is confirmed by examining the experimental results with $S$ set as small as one (i.e.~a single MCMC sample) on both the 20 newsgroups and TDT2 datasets, which are found to be very similar to the results with $S=10$ that are  shown in Figures~\ref{fig:20newsgroup}~and~\ref{fig:TDT2}. 
This is not surprising since it is not the absolute magnitude of the category-specific predictive likelihoods, only their relative rankings, that determine the categorization accuracy.

\begin{figure}[!tb]
\begin{center}
\includegraphics[width=0.85\columnwidth]{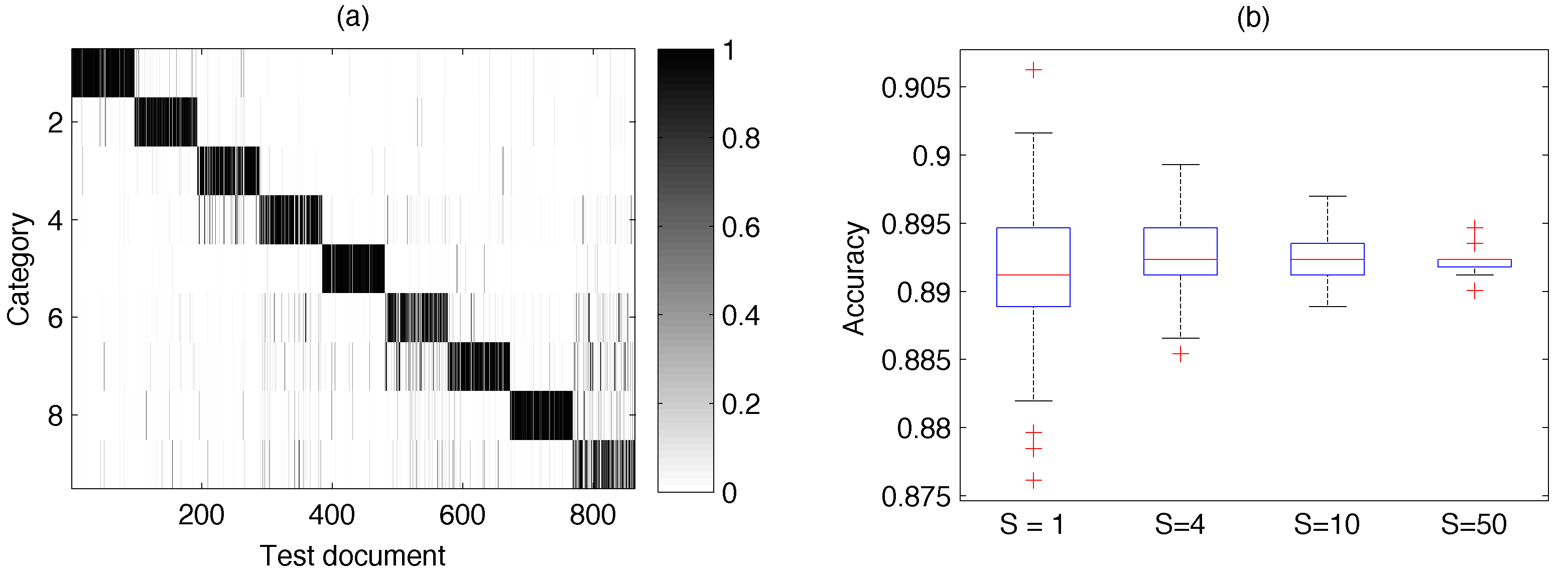}
\end{center}
\vspace{-3.5mm}
\caption{ \label{fig:S}  
(a) The predicted probabilities of the test documents under different categories for the CNAE-9 dataset, using the GNBP nonparametric Bayesian naive Bayes classifier with  20\% of the documents of each of the nine categories used for training. Each column shows the estimated probabilities across all nine categories for a singe document. Because the test documents from left to right were arranged from small to large according to their class labels,  
the dark diagonal band shows that most documents were placed with high posterior probability into the correct class.  (b) Monte Carlo variability of document categorization accuracies under different settings of $S$, the number of MCMC samples used in computing the predictive likelihood.  The boxplots show the variability of categorization accuracy when using $S=1$, $S=4$, $S=10$, and $S=50$ MCMC samples.  While the variability is clearly higher with fewer samples, there is no evident bias for using a small $S$, and the actual scale of the variability (standard error $<1\%$) is quite modest.}
\end{figure}

To further elaborate on this point, we consider the CNAE-9 dataset\footnote{\href{https://archive.ics.uci.edu/ml/datasets/CNAE-9}{https://archive.ics.uci.edu/ml/datasets/CNAE-9}} of \citet{ciarelli2009agglomeration}, which contains 1080 documents of free text business descriptions of Brazilian companies divided  into  nine categories, with a vocabulary size of $V=856$; and we randomly select
20\% of documents from each category as training, and calculate each test document's predictive probabilities  under the nine  categories, using the GNBP naive Bayes classifier with $S=1000$ samples, each of which is the 2500th MCMC sample of an independent Markov chain.  As shown in Figure \ref{fig:S} (a), in most cases, there is a little ambiguity on which category a test document should be assigned to. Hence letting $S=1000$ or $S=1$ make little practical difference in terms of categorization accuracy. In Figure \ref{fig:S} (b), from the left to  right,  we show the boxplot of 1000 accuracies produced by 1000 independent runs of the same testing procedure, each of which is calculated with $S=1$ MCMC sample; the boxplot of 250 accuracies with $S=4$; the boxplot of 100 accuracies with $S=10$; and the boxplot of 20 accuracies with $S=50$. It is clear from Figure \ref{fig:S} (b) that the larger the $S$ is, the less  the categorization accuracy varies, which is expected as the error of Monte Carlo integration decreases with $\sqrt{N}$. However, there is no substantial improvement for  the mean of the accuracies as   $S$ increases.  Even with $S=1$, the worst categorization accuracy is not too far from its mean. Therefore, in practice one may simply choose a small $S$ to compute the predictive likelihoods for the purpose of document categorization. 

  As opposed to the conventional multinomial naive-Bayes classifier that estimates the probability of each word in the vocabulary by normalizing the word counts, the proposed negative binomial processes provide new methods that directly  analyze the raw counts and take into account the total length of a document.  Moreover, there is no need to predetermine the vocabulary, as new features not present in the training data have been  taken care of by the nonparametric Bayesian predictive distributions of the negative binomial processes that are discussed in Section \ref{sec:predict_vector}.

\section{Conclusions}
This paper fills a gap in the nonparametric Bayesian literature,  deriving a family of probability mass functions for random count matrices by exploiting the gamma-Poisson, gamma-negative binomial, and beta-negative binomial processes.  
The resulting random count matrices have a random number of i.i.d.~columns, and their parameters can be inferred with closed-form update equations.  Any random count matrix in this family can be constructed by generating all its i.i.d. columns at once, or by adding one row at a time.  Our results also allow us to define the predictive distribution of an infinite-dimensional random count vector under any of the proposed priors, leading to three nonparametric Bayesian naive Bayes classifiers for count vectors.  The proposed classifiers, which directly operate on the raw counts and require no parameter tuning, alleviate the need to predetermine a shared finite vocabulary, and can account for features not present in the training data.  Example results on document categorization show that the proposed gamma-negative binomial process and beta-negative binomial process clearly outperform  both the negative binomial process and the multinomial naive Bayes classifier with Laplace smoothing, and have comparable performance to other state-of-the-art discriminatively-trained  text  classification algorithms. We are currently extending  
the techniques developed here 
to 
construct nonparametric Bayesian priors for a random count matrix, which has an unbounded number of columns and each row of which sums to a fixed integer; 
this extension can be used to construct nonparametric Bayesian discrete latent variable models, whose feature usages are represented with infinite random count matrices that are not directly observable. 

\section*{Acknowledgements} 

The authors thank the editor, associate editor, and two anonymous referees, whose invaluable comments and suggestions have helped us to improve the paper substantially. JGS acknowledges the support of CAREER grant DMS-1255187 from the U.S. National Science Foundation.

\end{spacing}

\begin{spacing}{1}
\begin{small}
\bibliography{References102014}
\bibliographystyle{plainnat}
\end{small}
\end{spacing}

\newpage

\appendix
\numberwithin{equation}{section}

\begin{center}\Large{
Priors for Random Count Matrices Derived from a Family of Negative Binomial Processes: Supplementary Material}
\end{center}

\begin{spacing}{1.5}

\section{The Negative Binomial Process: Details}\label{sec:NBP}

\subsection{Negative binomial process random count matrix}\label{sec:NBP_marginal}

 To generate a random count matrix, we 
construct a gamma-Poisson process as 
\beq\label{eq:NBP}
X_j\sim\mbox{PP}(G),~G\sim\Gamma\mbox{P}(G_0,1/c).
\eeq 
\citet{NBP2012} derives the marginal distribution of  $X=\sum_{j=1}^J X_j$ and calls it as the negative binomial process (NBP), a draw from which is represented as an exchangeable random count \emph{vector}.  We do not consider that simplification in this paper and consequently our definition of the NBP, a draw from which is represented as a row-column exchangeable random count \emph{matrix}, differs from the one in \citet{NBP2012}.

The conditional likelihood  in (\ref{eq:NBP_Like})  can be re-written as $$p(\{X_j\}_{1,J} \mid G) = e^{-JG(\Omega) }\prod_{k=1}^{K_J} \sum_{k'=1}^\infty \frac{r_{k'}^{n_{\cdotv k'}}}{ \prod_{j=1}^J n_{jk'}!} \delta(\omega_{k'} = \omega_k)\, .$$  
Applying the Palm formula \citep{daley1988introduction,james2002poisson,bertoin2006random,
CarTehMur2013a} to the expectation $\E_G[p(\{X_j\}_{1,J} \mid G)]$, we have
\begin{align}
&\E_G[p(\{X_j\}_{1,J} \mid G)] = \E\left[e^{-JG(\Omega)}\prod_{k=1}^{K_J} \sum_{k'=1}^\infty \frac{r_{k'}^{n_{\cdotv k'}}}{ \prod_{j=1}^J n_{jk'}!} \delta(\omega_{k'} = \omega_k)\right] \notag\\
& = \int_{\mathbb{R}_+\times \Omega} \frac{r_{1}^{n_{\cdotv 1}}}{ \prod_{j=1}^J n_{j1}!} e^{-Jr_1} \nu(dr_1 d\omega_1) \E\left[e^{-JG(\Omega\backslash\{\omega_1\}) }\prod_{k=2}^{K_J} \sum_{k'=1}^\infty \frac{r_{k'}^{n_{\cdotv k'}}}{ \prod_{j=1}^J n_{jk'}!} \delta(\omega_{k'} = \omega_k)\right]\notag\\
&=\ldots\notag\\
&=\left\{\prod_{k=1}^{K_J}\int_{\mathbb{R}_+\times \Omega}\frac{r_{k}^{n_{\cdotv k}}}{ \prod_{j=1}^J n_{jk}!} e^{-Jr_k} \nu(dr_k d\omega_k)\right\} \cdot \left\{\E_G\left[e^{-JG(\Omega\backslash\mathcal{D}_J)}\right]\right\}.\notag
\end{align}
Directly calculation with $\int_{\mathbb{R}_+\times \Omega} r^{n}e^{-Jr} \nu(drd\omega) = \gamma_0 (J+c)^{-n} \Gamma(n) $ and $\E_G[e^{-JG(\Omega\backslash\mathcal{D}_J) }]= (1+J/c)^{-\gamma_0}$ leads to
\beq
p(\{X_j\}_{1,J} \mid \gamma_0, c) = \E_G[p(\{X_j\}_{1,J} \mid G)] = \gamma_0^{K_J}e^{-\gamma_0\ln(\frac{J+c}{c})} \prod_{k=1}^{K_J} \frac{ \frac{\Gamma(n_{\cdotv k})}{(J+c)^{n_{\cdotv k}}}}{\prod_{j=1}^J n_{jk}!}.\notag
\eeq

\section{Gamma-Negative Binomial Process: Details}\label{sec:GNBP}

\subsection{GNBP random count matrix}

Given the gamma process $G \sim \Gamma{\mbox{P}}(G_0,1/c)$, we define $X\mid G\sim\mbox{NBP}(G,p)$ as a negative binomial process such that $X(A)\sim\mbox{NB}(G(A),p)$ for each $A\subset \Omega$.
Replacing the Poisson processes in (\ref{eq:NBP}) with the negative binomial processes defined in this way yields a gamma-negative binomial process (GNBP):
\beq
X_j \sim \mbox{NBP}(G, p_j) \, , \quad  G \sim \Gamma{\mbox{P}}(G_0,1/c) \, .\notag
\eeq
With a draw from the gamma process $G \sim \Gamma{\mbox{P}}(G_0,1/c)$ expressed as $G=\sum_{k=1}^\infty r_k\delta_{\omega_k}$,
a draw from $X_j \mid G \sim \mbox{NBP}(G, p_j)$ can be expressed as
$
X_j = \sum_{k=1}^\infty n_{jk}\delta_{\omega_k},~n_{jk}\sim\mbox{NB}(r_k,p_j). \notag
$  
%
%
The GNBP employs row-specific probability parameters $p_j$ to model row heterogeneity, and hence $X_j$ are conditionally independent but not identically distributed if  $p_j$  at different rows are set differently. 
Note that 
the GNBP is previously proposed in \citet{NBP2012}, which focuses on finding the conditional posterior of $G$, without considering the marginalization of~$G$. 

The GNBP hierarchical construction is conceptually simple, but to obtain a random count matrix, we have to marginalize out the gamma process $G\sim\Gamma{\mbox{P}}(G_0,1/c)$.  
 As it is difficult to directly marginalize $G$ out of the
conditional likelihood of the observed $J$ rows as
\begin{align}
p(\{X_j\}_{1,J} \mid G,\pv)  
&= \prod_{k=1}^\infty \prod_{j=1}^J \frac{\Gamma(n_{jk}+r_k)}{ n_{jk}!\Gamma(r_k)}p_j^{n_{jk}}(1-p_j)^{r_k}, \notag 
\end{align}
where $\pv:=(p_1,\ldots,p_J)$, we first augment each 
 $n_{jk}\sim\mbox{NB}(r_k,p_j)$ under its compound Poisson representation as 
$
 n_{jk}\sim\mbox{SumLog}(l_{jk},p_j) ,~l_{jk}\sim\mbox{Pois}(r_kq_j).\notag
$

Define $X\sim\mbox{SumLogP}(L,p)$ as a sum-logarithmic process  such that $X(A)\sim\mbox{SumLog}(L(A),p)$ for each $A\subset \Omega$. With $X_j\sim\mbox{NBP}(G,p_j)$ augmented  as $X_j\sim
\mbox{SumLogP}(L_j,p_j),~L_j\sim\mbox{PP}(q_jG)$,
we may express the joint likelihood of $X_j$ and $L_j$ as
\begin{align}
p(\{X_j,L_j\}_{1,J}\mid G,\pv) 
&=  \prod_{j=1}^J \prod_{k=1}^\infty    \frac{|s(n_{jk},l_{jk})|r_k^{l_{jk}}}{n_{jk}!} p_j^{n_{jk}} (1-p_j)^{r_k} ,\notag
\end{align}
With $l_{\cdotv k}:=\sum_{j=1}^J l_{jk}$, similar to the analysis in Section \ref{sec:NBP}, 
we can 
reexpress the likelihood as
\begin{align}\label{eq:marginalLike}
p(\{X_j,L_j\}_{1,J} \mid G,\pv) 
  &= {e^{-q_{\cdotv}G(\Omega\backslash \mathcal{D})}}\prod_{k=1}^{K_J}r_k^{l_{\cdotv k}} e^{-q_{\cdotv}r_k} \left( \prod_{j=1}^J\frac{ |s(n_{jk},l_{jk})|p_j^{n_{jk}}}{n_{jk}!}\right).
\end{align}

Similar to the analysis in Section \ref{sec:NBP_marginal}, 
with $G$ marginalized out as $p(\{X_j,L_j\}_{1,J}\mid \gamma_0,c,\pv) =  \E_G[ p(\{X_j,L_j\}_{1,J}\mid G,\pv) ]$, 
we obtain the GNBP random matrix prior in  (\ref{eq:marginal_gamma0}) using
\beq\label{eq:marginal_gamma0_1}
f(\Nmat_J,\Lmat_J\mid \gamma_0,c,\pv) = \frac{p(\{X_j,L_j\}_{1,J}\mid \gamma_0,c,\pv) }{K_J!}.
\eeq

Although not obvious, one may verify that (\ref{eq:marginal_gamma0}) defines the PMF of a compound random count matrix, which can be generated via
\begin{align}
&n_{jk} \sim \mbox{SumLog}(l_{jk},p_j),\nonumber\\
&(l_{1k},\ldots,l_{Jk}) \sim \mbox{Mult}( l_{\cdotv k}, {q_1}/{q_{\cdotv}},\ldots,{q_J}/{q_{\cdotv}}),\nonumber\\
&l_{\cdotv k} \sim \mbox{Log}[{q_{\cdotv}}/{(c+q_{\cdotv}})],\nonumber\\
 &K_J \sim \mbox{Pois}\{\gamma_0[\ln(c+q_{\cdotv})-\ln(c)]\}.\label{eq:GNBP_Matrix}
 \end{align}
Let $\sigma(1),\ldots, \sigma(J)$ denote a random permutation of the column indices. 
If 
$p_j$ are set differently for different rows, then  $\mbox{Mult}( l_{\cdotv k}, {q_{\sigma(1)}}/{q_{\cdotv}},\ldots,{q_{\sigma(J)}}/{q_{\cdotv}})  \buildrel d \over \neq  \mbox{Mult}( l_{\cdotv k}, {q_1}/{q_{\cdotv}},\ldots,{q_J}/{q_{\cdotv}}) $ and hence   the introduced random count matrix no longer maintains row exchangeability. 

 Comparing (\ref{eq:GNBP_Matrix}) with (\ref{eq:NBProwwise}), one may identify several key differences between the GNBP and NBP random count matrices. First, one may increase $p_j$ to encourage the $j$th row to have larger counts than the others.  Second, both $n_{jk}$ and the column sum $n_{\cdotv k}$ are generated from compound distributions. In fact, 
  if we let $p_j\equiv 1-e^{-1}$, then the matrix $\{l_{jk}\}_{jk}$ in (\ref{eq:GNBP_Matrix}) is exactly a NBP random count matrix, and the GNBP builds its random matrix using $n_{jk} \sim \mbox{SumLog}(l_{jk},p_j) 
 $.

The sequential construction of a GNBP random count matrix can be intuitively explained as drawing dishes, drawing tables at each dish, and then drawing customers at each table. 
Similar to the definition of $\Nmat^{+}_{J+1}$,  
we let $\Lmat^{+}_{J+1}$ represent the new row and columns added to~$\Lmat_J$. 
Using (\ref{eq:marginal_gamma0}), following the analysis in Section~\ref{sec:NBP0}, one may show with direct calculation~that
\begin{align}\label{eq:GNBP_pre}
p(\Nmat^{+}_{J+1}, \Lmat^{+}_{J+1}\mid \Nmat_J, \Lmat_J,\thetav)
&=\frac{K_J! K^+_{J+1}!}{K_{J+1}!}\prod_{k=1}^{K_{J+1}}\mbox{SumLog}\left(l_{(J+1)k},p_{J+1}\right)\notag\\
&\times\prod_{k=1}^{K_J} \mbox{NB}\left(l_{(J+1)k};l_{\cdotv k}, \frac{q_{J+1}}{c+q_{\cdotv} + q_{J+1}}\right)\notag\\
&\times\prod_{k=K_J+1}^{K_{J+1}} \mbox{Log}\left(l_{(J+1)k}; \frac{q_{J+1}}{c+q_{\cdotv} + q_{J+1}}\right)\notag\\
&\times\mbox{Pois}\left\{K^+_{J+1};\gamma_0 \left[\ln(c+q_{\cdotv}+q_{J+1}) - \ln(c+q_{\cdotv}) \right]\right\}.
\end{align}
Thus to add a new row,  we first draw
$
\mbox{NB}[l_{\cdotv k}, {q_{J+1}} / {(c+q_{\cdotv} + q_{J+1}) }]
$
tables at existing columns (dishes); we then draw 
$
K^+_{J+1}\sim\mbox{Pois}\{\gamma_0[\ln(c+q_{\cdotv}+q_{J+1})-\ln(c+q_{\cdotv})]\} 
$
new dishes, each of which is associated with
$
\mbox{Log}[{q_{J+1}} / {(c+q_{\cdotv} + q_{J+1})}]
$
tables; we further draw $\mbox{Log}(p_{J+1})$ customers at each table  and aggregate the counts across the tables of the same dish
as
$
n_{(J+1)k} = \sum_{t=1}^{l_{(J+1)k}} n_{(J+1)kt};
$
and in the final step, we insert the $K^+_{J+1}$ new columns into the $K_{J}$ original columns without reordering, which again is a one to $K_{J+1}!/\left(K_J! \ K^+_{J+1}!\right)$ mapping.
We emphasize that the number of tables (customers) for a new dish, which follows a logarithmic (sum-logarithmic) distribution, must be at least one; the implication is that there are infinite many dishes  that have not yet been ordered by any of the tables seated by existing customers.
The sequential construction provides a convenient way to construct a GNBP random count matrix one row at a time.

With the latent counts 
$l_{(J+1)k}$ marginalized out, one may show that the predictive distribution for $\Nmat^{+}_{J+1}$, given $\Nmat_J$ and $\Lmat_J$, can be expressed in terms of the Poisson, LogLog and GNB distributions as
\begin{align}\label{eq:GNBP_pre1}
p(\Nmat^{+}_{J+1} \mid \Nmat_J, \Lmat_J,\thetav)
&=\frac{K_J! K^+_{J+1}!}{K_{J+1}!} \prod_{k=1}^{K_J} \mbox{GNB}\left(n_{(J+1)k};l_{\cdotv k},c+q_{\cdotv} , p_{J+1} \right)\notag\\
&\times\prod_{k=K_J+1}^{K_{J+1}} \mbox{LogLog}\left(n_{(J+1)k};c+q_{\cdotv} , p_{J+1} \right)\notag\\
&\times\mbox{Pois}\left\{K^+_{J+1};\gamma_0 \left[\ln(c+q_{\cdotv}+q_{J+1}) - \ln(c+q_{\cdotv}) \right]\right\}, 
\end{align}
where $n\sim\mbox{LogLog}(c,p)$ represents a logarithmic mixed sum-logarithmic distribution defined on positive integers  and $n\sim\mbox{GNB}(l,c,p)$ represents a gamma mixed negative binomial distribution defined on $\mathbb{Z}$, whose PMFs are shown in Appendix \ref{sec:dist}.


\subsection{Inference for parameters}
Both the GNB and LogLog distributions have complicated PMFs involving Stirling numbers of the first kind and it seems difficult to infer their parameters. 
Fortunately, using the likelihoods (\ref{eq:marginalLike}) and (\ref{eq:marginal_gamma0}) and the data augmentation techniques  developed for the negative binomial distribution \citep{NBP2012}, we are able to derive closed-form conditional posteriors for the GNBP. To complete the model, we let  $\gamma_0\sim \mbox{Gamma}(e_0, {1}/{f_0})$, $p_j\sim\mbox{Beta}(a_0,b_0)$ and $c\sim\mbox{Gamma}(c_0,1/d_0)$. We sample the model parameters as
\begin{align}\label{eq:GNBP_sampling}
&(\gamma_0|-)\sim\mbox{Gamma}\bigg(e_0+K_J, \frac{1}{f_0-\ln(\frac{c}{c+q_{\cdotv}})}\bigg),\notag\\
&(l_{jk}|-)=\sum_{t=1}^{n_{jk}}u_t,~u_t\sim\mbox{Bernoulli}\bigg(\frac{r_k}{r_k+t-1}\bigg),\notag\\
&(r_k|-)\sim\mbox{Gamma}\big( l_{\cdotv k}, 1/(c+q_{\cdotv})\big),\notag\\ 
&\{G(\Omega\backslash\mathcal{D}_J)|-\}\sim\mbox{Gamma}\big(\gamma_0,1/(c+q_{\cdotv})\big),\notag\\ 
&(p_j|-)\sim\mbox{Beta}\big(a_0+m_j,b_0 + G(\Omega)\big),\notag\\ 
&(c|-)\sim\mbox{Gamma}\big(c_0+\gamma_0,{1}/{[d_0 + G(\Omega)]\big)}. 
\end{align}

\section{Beta-Negative Binomial Process: Details}\label{sec:BNBP}
\subsection{BNBP random count matrix}

The GNBP generalizes the NBP by replacing the Poisson process in (\ref{eq:NBP}) using a negative binomial process and shares the negative binomial dispersion parameters across rows. 
Exploiting an alternative strategy that shares the negative binomial probability parameters  across rows, 
we  
construct a BNBP as
\beq
X_j\sim\mbox{NBP}(r_j, B),~B\sim\mbox{BP}(c,B_0),\notag
\eeq
where $p_k = B(\omega_k)$ is the weight of the atom $\omega_k$ of the beta process $B\sim\mbox{BP}(c,B_0)$, and $X_j \mid B\sim\mbox{NBP}(r_j,B)$ is a negative binomial process such that $X_j(A)=\sum_{k:\omega_k\in A} n_{jk},~n_{jk}\sim\mbox{NB}(r_j,p_k)$ for each $A\subset \Omega$.

With $\rv:=(r_1,\ldots,r_J)$, similar to the analysis in Appendix~\ref{sec:GNBP}, the  likelihood of the BNBP can be expressed as
\begin{align}\label{eq:Likelihood_BNBP1}
p(\{X\}_{1,J}\mid B,\rv)=e^{-p_* r_{\cdotv}}
\prod_{k=1}^{K_J}  p_k^{n_{\cdotv k}}(1-p_k)^{r_{\cdotv}} \prod_{j=1}^J \frac{\Gamma(n_{jk}+r_j)}{n_{jk}!\Gamma(r_j)},
\end{align}
where $p_*$ denotes the sum over all the atoms in the absolutely continuous space $\Omega\backslash\mathcal{D}_J$ as
\beqs
&p_* := - \sum_{k:n_{\cdotv k}=0} \ln(1-p_k)\notag
\eeqs
and $r_{\cdotv}:=\sum_{j=1}^J r_j$.
Using the  L\'evy-Khintchine theorem and (\ref{eq:BP_Levy}), the Laplace transform of~$p_*$ can be expressed as
\begin{align}\label{eq:LaplaceP}
\E[e^{-s p_*}] 
&=\exp\left\{{\int_{[0,1]\times\Omega} \left[(1-p)^{s}-1\right]\nu(dpd\omega) }\right\}\notag\\
&=\exp\left[{-\gamma_0\sum_{i=0}^\infty \left(\frac{1}{c+i}-\frac{1}{c+i+s} \right)}\right]\notag\\
&=\exp\left\{{-\gamma_0\left[\psi(c+s)-\psi(c)\right]}\right\},\notag
\end{align}
where $\psi(x)={\Gamma'(x)}/{\Gamma(x)}$ is the digamma function; we define such a random variable as the logbeta random variable
\beq
p_*\sim\mbox{logBeta}(\gamma_0,c),\notag
\eeq
whose mean and variance are $\E[p_*]=\gamma_0\psi_1(c)$ and $\mbox{Var}[p_*]=-\gamma_0\psi_2(c)$, respectively, where $\psi_{n}(x)=\frac{d^n \psi(x)}{dx^n}$.

As before, one may verify with direct calculation that (\ref{eq:BNBP_Matrix}) defines
the PMF of a column-i.i.d. random count matrix $\Nmat_J\in\mathbb{Z}^{ J\times K_J}$, which  can be generated via
\begin{align}\label{eq:BNBP_Matrix1}
&\nv_{:k}\sim\mbox{DirMult}(n_{\cdotv k}, r_1,\ldots,r_J),\notag\\
&n_{\cdotv k} \sim \mbox{Digam}(r_{\cdotv}, c),\notag\\
&K_J\sim  \mbox{Pois}\big\{ \gamma_0\left[\psi(c+r_{\cdotv})-\psi(c)\right]\big\},
 \end{align}
where the PMFs of both the Dirichlet-multinomial (DirMult) and digamma distributions  are shown in the Appendix. 
Note that if
$r_j$ are set differently for different rows, then
$
\mbox{DirMult}(n_{\cdotv k}, r_{\sigma(1)},\ldots,r_{\sigma(J)}) \buildrel d \over \neq \mbox{DirMult}(n_{\cdotv k}, r_1,\ldots,r_J)
$
and hence   the corresponding random count matrix no longer maintains row exchangeability.

The sequential construction of a BNBP random count matrix 
can be intuitively understood as an ``ice cream'' buffet process (ICBP). 
Using (\ref{eq:BNBP_Matrix}), similar to the analysis in Section \ref{sec:NBP0}, we have
\begin{align} \label{eq:BNBP_pre}
p(\Nmat^{+}_{J+1} \mid \Nmat_J)
&=\frac{K_J! K^+_{J+1}!}{K_{J+1}!} \prod_{k=1}^{K_J} \mbox{BNB}(n_{(J+1)k};r_{J+1},n_{\cdotv k}, c+r_{\cdotv})\notag\\
&\times\prod_{k=K_J+1}^{K_{J+1}} \mbox{Digam}(n_{(J+1)k};r_{J+1}, c+r_{\cdotv})\notag\\
&\times\mbox{Pois}\left\{K^+_{J+1};\gamma_0\left[\psi(c+r_{\cdotv}+r_{J+1})-\psi(c+r_{\cdotv})\right]\right\}, 
\end{align}
where the PMF for the beta-negative binomial (BNB) distribution is shown in Appendix \ref{sec:dist}. 
Thus to add a row to $\Nmat_J\in\mathbb{Z}^{J \times K_J}$, 
customer $J+1$ takes 
$ 
n_{(J+1)k}
\sim\mbox{BNB}(r_{J+1},n_{\cdotv k}, c+r_{\cdotv}) 
$ 
number of scoops at an existing ice cream (column); 
the customer further  selects
$ 
K^+_{J+1}\sim\mbox{Pois}\left\{\gamma_0\left[\psi(c+r_{\cdotv}+r_{J+1})-\psi(c+r_{\cdotv})\right]\right\}
$ 
new ice creams out of the buffet line and takes
$ 
n_{(J+1)k}\sim\mbox{Digam}(r_{J+1}, c+r_{\cdotv})
$ 
number of scoops at each new ice cream. Thus the ICBP can also be considered as a ``multiple-scoop'' Indian buffet process, an analogy used in \citet{BNBP_PFA_AISTATS2012}.
Note that when $r_j\equiv 1$, we have $K^+_{J+1}\sim\mbox{Pois}[{\gamma_0 }/{(c+J)}]$, confirming the derivation about the number of new dishes (ice creams) in Section 3.2 of \citet{BNBP_PFA_AISTATS2012}\footnote{Due to different parameterization of the L\'evy measure, the beta process mass parameter $\gamma_0$ in this paper can be considered as $\gamma_0 c$ in \citet{JordanBP} and \citet{BNBP_PFA_AISTATS2012}.}, which, however, provides no descriptions about the distributions of the number of scoops at existing and new ice creams. We emphasize that the number of scoops at a new ice cream, which follows a digamma distribution, must be at least one; the implication is that there are infinite many ice creams in the buffet line that have not yet been scooped by any of the existing customers. 
Similar to the GNBP random count matrix, the BNBP random count matrix is column exchangeable, but not row exchangeable if the row-specific dispersion parameters $r_j$ are fixed at different values. 

A related marked BNBP of \citet{BNBP_PFA_AISTATS2012,NBP_NIPS2012} attaches  an independent negative binomial dispersion parameter $r_k$ for each atom of the beta process, and infers its values under a finite approximation of the beta process; another related BNBP of \citet{NBPJordan} uses a single  dispersion parameter $r$ and sets its value empirically. 
None of these papers, however, marginalize out the beta process to define a prior on column-i.i.d. random count matrices, a challenge tackled in this paper.

 Independently of our work, \citet{NBIBP} also describe the marginalization of the beta process from the negative binomial process, where the obtained BNBP is called the negative binomial Indian buffet process. Although the idea of marginalizing out the beta process is shared by both papers, the techniques and combinatorial arguments used are quite different. 
Their paper focuses on a special case of the BNBP where a single dispersion parameter $r$ is used for all the $X_j$'s.  
Our model allows row-specific dispersion parameters  $r_j$, develops an efficient inference scheme  for all model parameters, derives the predictive distribution of a new row count vector under a BNBP random count matrix, and also situates the BNBP in the larger family of count-matrix priors derived from negative-binomial processes.

\subsection{Inference for parameters}\label{sec:BNBP_sampling}
For all the  atoms in the absolutely continuous part of the space, $\Omega\backslash\mathcal{D}_J$, we have that 
$$
(\nu(dpd\omega)\mid -) = p^{-1}(1-p)^{c+r_{\cdotv}-1}dp B_0(d\omega) \, .
$$ 
Thus the Laplace transform of $(p_*|-)$ 
can be expressed as
\begin{align}
\E[e^{-s (p_*|-)}]  
&=\exp \left\{-\gamma_0\left[\psi(c+r_{\cdotv}+s)-\psi(c+r_{\cdotv})\right] \right\},\notag
\end{align}
and hence we have
$
(p_*|-)\sim\mbox{logBeta}(\gamma_0 ,c+r_{\cdotv}).
$
With its Laplace transform, we sample $(p_*|-)$ using the method proposed in \citet{ridout2009generating}. 
To complete the model, we let  $\gamma_0\sim \mbox{Gamma}(e_0, {1}/{f_0})$, $r_j\sim\mbox{Gamma}(a_0,b_0)$ and $c\sim\mbox{Gamma}(c_0,1/d_0)$. 
Using both the conditional likelihood (\ref{eq:Likelihood_BNBP1}) and the marginal likelihood (\ref{eq:BNBP_Matrix}),  and the data augmentation techniques developed 
in \citet{NBP2012}, we sample the model parameters as
\begin{align}\label{eq:BNBP_sampling}
&(\gamma_0|-)\sim\mbox{Gamma}\bigg(e_0+K_J,\frac{1}{f_0+\psi(c+r_{\cdotv})-\psi(c)}\bigg),\notag\\
&(p_k|-)\sim \mbox{Beta}(n_{\cdotv k},c+r_{\cdotv}),~~(p_*|-)\sim\mbox{logBeta}(\gamma_0 ,c+r_{\cdotv}), \notag\\ 
&(l_{jk}| - ) 
=\sum_{t=1}^{n_{jk}}u_t,~u_t\sim\mbox{Bernoulli}\bigg(\frac{r_j}{r_j+t-1}\bigg),\notag\\
&(r_j|-)\sim\mbox{Gamma}\bigg(a_0 + l_{j\cdotv} ,\frac{1}{b_0+ p_*-\sum_{k=1}^{K_J}\ln(1-p_k)} \bigg).
\end{align}
The only parameter that does not have an analytic conditional posterior is the concentration parameter $c$. 
Since using Campbell's theorem \citep{PoissonP}, we have  $\E[\sum_{k} p_k] = \int_{[0,1]\times \Omega} p \nu(dpd\omega) = \gamma_0/c$, 
to sample  $c$, we use 
\beq\label{eq:BNBP_sampling1}
Q(c')=\mbox{Gamma}\bigg(c_0+\gamma_0,\frac{1}{d_0+p_*+\sum_{k=1}^{K_J} p_k }\bigg)
\eeq
as the proposal distribution in an independence chain Metropolis-Hastings sampling step. 
One may also sample $c$ using a griddy-Gibbs sampler \citep{griddygibbs}.

\section{Some useful distributions}\label{sec:dist}

Direct calculation shows that the logarithmic mixed sum-logarithmic (LogLog) distribution, expressed as $
n\sim\mbox{SumLog} (l,p),~l\sim\mbox{Log}\Big(\frac{-\ln(1-p)}{c-\ln(1-p)}\Big),\notag
$
 has PMF
\begin{align}
f_N(n|c,p) 
& = \frac{ \sum_{l=1}^n \frac{ |s(n,l)|p^n}{n!} \frac{\Gamma(l)}{[c-\ln(1-p)]^{l}}}{\ln[c-\ln(1-p)]-\ln(c)}\notag
\end{align}
for $n\in\{1,2,\ldots\}$;
and the negative binomial mixed sum-logarithmic distribution, expressed as 
$
n\sim\mbox{SumLog}(l,p),~l\sim\mbox{NB}\left(e,\frac{-\ln(1-p)}{c-\ln(1-p)}\right),
$
has PMF
\begin{align}
f_N(n|e,c,p)  
& = \sum_{l=0}^n \frac{c^{e}p^n|s(n,l)|}{\Gamma(e)n!}  \frac{\Gamma(e+l)}{[c-\ln(1-p)]^{e+l}}\notag
\end{align}
for $n\in\{0,1,\ldots\}$. The iterative calculation of ${ |s(n,l)|}/{n!}$ under the logarithmic scale is described in Appendix \ref{sec:Striling}. 
Using (\ref{eq:gammafunctions}), one may show that
the negative binomial mixed sum-logarithmic distribution shown above is equivalent to a gamma mixed negative binomial (GNB) distribution, generated by $n\sim\mbox{NB}(r,p),~r\sim\mbox{Gamma}(e,1/c)$. 
Note that $n\sim\mbox{LogLog}(c,p)$ is the limit of $n\sim\mbox{GNB}(e,c,p)$ as $e\rightarrow 0$, conditioning on $n>0$, thus it can be considered as a truncated GNB distribution.

The Dirichlet-multinomial (DirMult) distribution \citep{mosimann1962compound, 
madsen2005modeling} is a Dirichlet mixed multinomial distribution, with PMF
\beq
\mbox{DirMult}(\nv_{:k}\mid n_{\cdotv k}, \rv) =  \frac{n_{\cdotv k}!}{\prod_{j=1}^Jn_{kj}!} \frac{\Gamma(r_{\cdotv})}{\Gamma(n_{\cdotv k}+r_{\cdotv})} {\prod_{j=1}^J  \frac{\Gamma(n_{kj}+r_j)}{\Gamma(r_j)}},\notag
\eeq
and the digamma distribution \citep{digamma} has PMF
\beq\label{eq:digammaPMF}
\mbox{Digam}(n\mid r,c) = \frac{1 }{\psi(c+r)-\psi(c)}   \frac{\Gamma(r+n)\Gamma(c+r)}{n\Gamma(c+n+r)\Gamma(r)},
\eeq
where $n=1,2,\ldots$. 
 Since the 
beta-negative binomial (BNB) distribution has PMF
\begin{align}
f_N(n\mid r,e,c)&=\int_0^1 \mbox{NB}(n;r,p)\mbox{Beta}(p;e,c)dp=\frac{\Gamma(r+n)}{n!\Gamma(r)}\frac{\Gamma(c+r)\Gamma(e+n)\Gamma(e+c)}{\Gamma(e+c+r+n)\Gamma(e)\Gamma(c)},\notag
\end{align}
one may show that conditioning on $n>0$, $n\sim\mbox{BNB}(r,e,c)$ becomes $n\sim\mbox{Digam}(r,c)$ as $e\rightarrow 0$. Thus the digamma distribution can be considered as a truncated BNB distribution. 

Since the Laplace transform of the logbeta random variable $p_*\sim\mbox{logBeta}(\gamma_0,c)$ can  be reexpressed as
$$
\E[e^{-s p_*}] = \prod_{i=0}^\infty \exp\left\{{\frac{\gamma_0}{ c+i}\left[\left(1+\frac{s}{c+i}\right)^{-1} -1\right]}\right\},
$$
we can generate $p_*\sim\mbox{logBeta}(\gamma_0,c)$  as an infinite sum of independent compound Poisson random variables as
\begin{align}\label{eq:infitSum}
&p_*=\sum_{i=0}^\infty \lambda_i, ~\lambda_i=\sum_{t=1}^{u_i}\lambda_{it},~u_i\sim\mbox{Pois}\left(\frac{\gamma_0}{c+i}\right),~\lambda_{it}\sim\mbox{Gamma}\left(1,\frac{1}{c+i}\right).
\end{align}

\section{Calculating Stirling Numbers of the First Kind}\label{sec:Striling}
The unsigned Stirling numbers of the first kind $|s(n,l)|$ appear in the predictive distribution for the GNBP. 
It is numerically unstable to recursively calculate $|s(n,l)|$ based on $|s(n,l)| = (n-1)|s(n-1,l)| + |s(n-1,l-1)|$, as $|s(n,l)|$ would rapidly reach the maximum value allowed by a finite precision machine as $n$ increases. Denoting $$g(n,l)=\ln(|s(n,l)|) - \ln({n!}),$$
we iteratively calculate $g(n,l)$ with $g(n,1) =\ln(n-1)-\ln({n}) + \ln g(n-1,1)$, $g(n,n) =g(n-1,n-1) -\ln n$, and 
\begin{align}
g(n,l) &= \ln\frac{n-1}{n} +g(n-1,l) + \ln\left\{1+\exp[{g(n-1,l-1) - g(n-1,l) - \ln(n-1)}]\right\}\notag
\end{align}
for  $2\le l\le n-1$. This approach is found to be numerically stable.


\end{spacing}

\end{document}